# Electron-Light Interactions beyond the Adiabatic Approximation: Recoil Engineering and Spectral Interferometry

Nahid Talebi

*Max Planck Institute for Solid State Research, Stuttgart, Germany*

Email: n.talebi@fkf.mpg.de

Nahid Talebi is a scientist whose work focuses on the fields of numerical electrodynamics, electron-light interactions, plasmonics, and nanophotonics.

# Electron-Light Interactions beyond the Adiabatic Approximation: Recoil Engineering and Spectral Interferometry

The adiabatic approximation has formed the basis for much of our understandings of the interaction of light and electrons. The classical non-recoil approximation or quantum mechanical Wolkow states of free – electron waves have been routinely employed to interpret the outcomes of low-loss EELS or electron holography. Despite the enormous success of semianalytical approximations, there are certainly ranges of electron-photon coupling strengths where more demanding self-consistent analyses are to be exploited to thoroughly grasp our experimental results. Slow-electron point-projection microscopes and many of the photoemission experiments are employed within such ranges. Here we aim to classify those regimes and propose numerical solutions for an accurate simulation model. A survey of the works carried out within self-consistent Maxwell-Lorentz and Maxwell-Schrödinger frameworks are outlined. Several applications of the proposed frameworks are discussed, and an outlook for further investigations is also delivered.

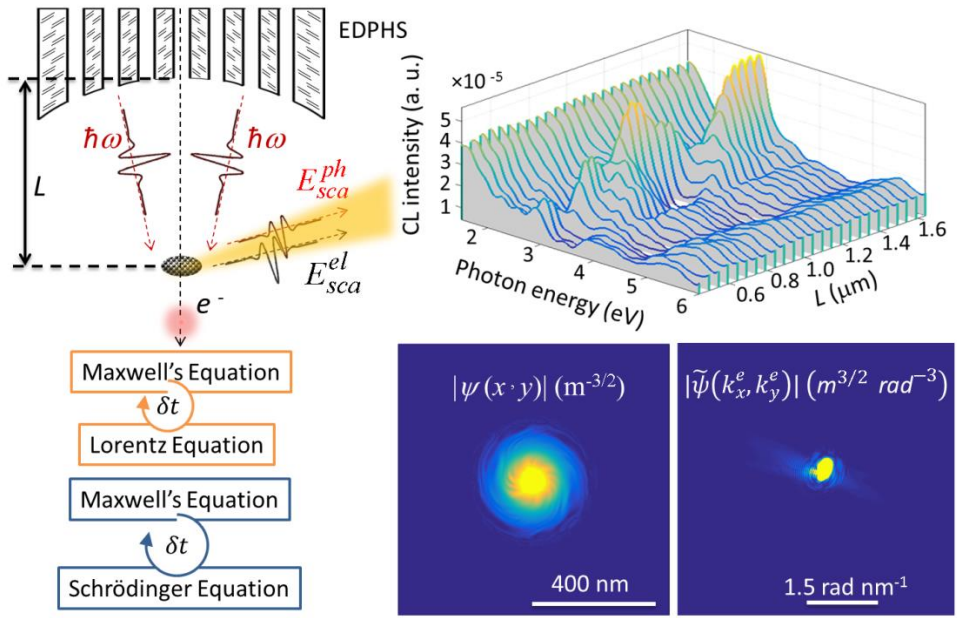

## 1. Introduction

Interaction of electron waves with light and matter has been a subject of intense studies in the last decades. This research area is reached from the point of view of first principles, due to the interesting quantum phenomena which it covers, like the Kapitza-Dirac effect [1] and Compton scattering [2], but also from practical aspects. Additionally, the numerous mechanisms of radiation of free-electron beams, from classical Larmor radiation to coherent x-ray radiation, have opened a new avenue for the design of modern light sources with collimated and coherent radiation properties [3-7]. Additionally, combining laser and electron guns in electron microscopes have created a plethora of opportunities in characterizing the chemical reactions and near-field distributions of nanostructures [8-15]. This method, which is called ultrafast electron diffraction or photon-induced near-field electron microscopy (PINEM), where for the former diffraction patterns and for the latter spectra are acquired, has been recently further developed by several groups around the world [16-20], into a form of time-resolved pump-probe characterization methodology. Photoemission electron guns are nowadays controlled to create sub-picosecond electron pulses with great spatial coherence, almost at the same level of field-driven electron guns [21]. Moreover, by controlling the laser-electron jitter by means of microwave or THz cavities, the longitudinal broadening of electron pulses has been considerably reduced [22-27]. Additionally, it was recently demonstrated that the interaction of electron pulses with laser-induced near-field of nanostructures will cause attosecond bunching of single-electron pulses in space-time [28].

Considering however the main purpose of any kind of characterization technique, what is most often demanded is to extract as much data as possible from the system under investigation. Especially imaging the dynamics of electrons in atoms or solid state systems is one of the key goals of ultrafast science. Along with this line, improving the spatiotemporal resolution of our methods towards attosecond time resolution and sub-Angstrom spatial resolution is highly required. However, within the context of PINEM and ultrafast electron diffraction, reaching attosecond time resolution seems to be challenging mostly due to the difficulty in the synchronization between electron and photon pulses [29-31]. Additionally, merely using the approaches stated above, electron-beam characterization methods might be able to coherently control physical and chemical processes in the samples, analogous to quantum coherent control with shaped light [32, 33], albeit with better spatial resolution. Another method which is based on an inverse approach to the photoemission processes, i.e. control of electron-based radiation mechanisms to create desired photon pulses, has been recently proposed [34, 35]. This approach can be used to retrieve the spectral phase, and in principle is able to push electron-based characterization techniques into the attosecond era.

All the above-stated approaches routinely employ relativistic electron beams, either in a pulsed or a continuous configuration. Besides the aforementioned characterization techniques, point-projection electron microscopy (PPM) has appeared as a compact and efficient imaging method, in its pioneering DC configuration developed by Fink and co-workers [36]. Slow-electrons (with kinetic energies below 1keV) appear to be a more sensitive probe of the electromagnetic field, and are much easier to shape and manipulate, in comparison with their relativistic counterparts. Low-energy electrons are also more practical probes of biological samples, which impose less radiation damage in comparison with higher energy electrons [37].

Additionally, PPM has been advanced by making use of metallic nanotips illuminated by short laser pulses, to form near-point femtosecond (fs) electron probes [38, 39]. Making use of adiabatic nanofocusing [40] it is possible to realize practical fs-PPM, where the sample can be placed in a close proximity of the nanotip electron gun. This in particular has advantages in retaining the electron pulses as short as demanded, by simply restricting the time-of-flight of electron pulses from the nanotip to the sample, hence limiting the pulse broadening due to the space-charge effect and achromatic aberrations.

Aligned with the technological developments of electron-based characterization techniques, our theoretical frameworks are yet to be adapted to the strong-laser and slow-electron regimes. More specifically, there exist certain domains where our adiabatic approximations might break down. This is practically important from several viewpoints: (i) in PPM the shape and amplitude of electron beams are both strongly manipulated, in addition to their phase, (ii) even in free-space electron-light interaction, purely elastic approximations might appear to be a mere over-simplification [41], (iii) during the interaction of electron beams with gratings and light, electron bunching appears to be an additional mechanism to the electron acceleration, where both acceleration and bunching mechanisms are controlled by the longitudinal broadening of the electron beam relative to the grating period [42], (iv) shaped electron beams interacting with matter have different selection rules and might offer approaches for manipulating the electron-induced radiations [43-45]. The latter point is fundamentally important, as even for a single electron wavepacket, when the electron beam is in a superposition of at least two momentum states, interferences between different quantum paths in interaction of photons with the electron may occur [46]. As noted by Keitel and co-workers, the quantum eigenstates of electrons in a nonplanar laser beam or in general shaped light waves are however unknown [46].

For this reason, the development of self-consistent numerical methods may facilitate a better understanding of the outcomes of experiments [34, 42, 47, 48] and simulate the design of new experiments.

The present work provides an overview of electron-light interaction, and the methodologies developed within recent years to control those interactions, from a new perspective, i.e., non-adiabatic analysis and recoil engineering. We address the problems using both classical and quantum-mechanical approaches. Starting from the electron-induced domain, we outline electron-based characterization methods which are used to understand the photonic local density of states in nanostructures. We thereafter describe the recoil that the electron receives in interaction with near-field distributions of nanostructures. In particular engineering principles to enhance the electron recoil will be addressed. Within this concept we review the advances in recently proposed methodologies based on spectral interferometry with electron microscopes. Thereafter, we briefly describe another point of view, i.e., adiabatic assumptions in quantum mechanics, to model the electron-light interaction. We further review self-consistent techniques and recent advances in numerical methods, with applications in understanding the outcomes of ultrafast PPM, of the photoemission process, and of linear accelerators.

## 2. Electron - induced domain

Electrons and photons are elementary particles whose interactions underline our understanding of the physical and chemical processes in samples. It is hence quite natural to consider electrons and photons as individual *incident* beams to initiate the excitation, and also to choose either scattered photons or electrons as *detection probes*. In this way our characterization methodologies can be divided into those groups which use either electrons or photons as incident beams. Here, we provide an overview of the former group, i.e., electron-beam characterization techniques where the electron beams are used to trigger the sample response.

Electron energy-loss spectroscopy (EELS) is an operational mode of transmission electron microscopes (TEMs), for which the amount of energy loss in interaction of electron beams with samples is detected. Electron energy-loss spectra are usually divided into two domains, namely low-loss and core-loss domains. The resonances in the low-loss domain mainly originate from collective excitations of valence and conduction electrons [49]. In contrast, it is single-electron inner-shell excitations and transitions cause the resonances in the core-loss domain. In this work, we only provide a brief review of the advances by low-loss EELS. A complete review of the field has been provided by Garcia de Abajo [50]; here we mainly address more recent works. For core-loss electron energy-loss spectra and fine spectral structures, the readers are referred to the works by Egerton [49] and Keast et al. [51].

Another probe of collective electron excitations is Cathodoluminescence (CL). Electron beams interacting with nanostructures emit light, which can be detected using a CL detectors. CL has been historically introduced for characterizing semiconductors, ceramic, and minerals [52]. Recent advances in CL however facilitate characterization of the electron-induced luminescence in a rather wide energy range from UV to IR [53-56], as well as understanding the coherent versus the incoherent nature of the radiation [57].

*2.1 Electron-mapping of collective excitations in nanostructures*

EELS and CL have been introduced as efficient tools for probing nanooptical excitations at single nanostructures, with nanometer spatial resolution and meV energy resolution. Thanks to the ultrafast interaction of localized relativistic electrons with the optical modes of nanostructures in TEMs, electron beams appear as an ultra-broadband probe of sample resonances. The inelastic interaction of a swift electron with nanostructures can be understood using a useful classical approach [50], which has been proven to be identical to the quantum-mechanical treatment when averaging over electron impact parameters weighted by the spot intensity [58]. Interpreting electron energy-loss spectra as the probability of the electron to lose an amount of energy equal to $\hbar\omega$, the loss-probability is given as

$$\Gamma^{\text{EELS}}(\omega) = \frac{e}{\pi\hbar\omega} \int dt \, \text{Re}\left\{ e^{-i\omega t} \vec{v}_e \cdot \vec{E}^{ind}\left[\vec{r}_e(t),\omega\right] \right\} \quad (1)$$

where $\vec{v}_e$ is the electron velocity, $\vec{r}_e(t)$ is the electron trajectory, $e$ is the electron charge, and $\hbar$ is the reduced Planck constant [50]. A complete treatment then requires the knowledge about the electron trajectory, as well as the induced electric field ($\vec{E}^{ind}$) along the electron trajectory. Assuming a uniform electron trajectory along the *z*-axis as $\vec{r}_e(t) = (x_0, y_0, z = v_e t)$ (nonrecoil approximation), Eq. (1) can be further simplified as

$$\begin{aligned}\Gamma^{\text{EELS}}(x_0, y_0, \omega) &= \frac{e}{\pi\hbar\omega} \text{Re} \int_{-\infty}^{+\infty} dz \, \tilde{E}_z(x_0, y_0, z; \omega) \, e^{i\frac{\omega}{v_e}z} \\ &= \frac{e}{\pi\hbar\omega} \text{Re} \, \tilde{E}_z(x_0, y_0, k_z = \omega/v_e; \omega)\end{aligned} \quad (2)$$

which is clearly signifying the role of momentum conservation during the interaction of electrons with near-field distributions of nanostructures as $\hbar k_z = \hbar\omega/v_e$. Interestingly, the loss probability is directly related to the induced-electric field, which renders EELS a powerful technique to directly map the electric field projected along the electron trajectory [59]. One might find another useful derivation of the loss probability by treating EELS as the *rate of energy leaving the electron beam* as [60]

$$\Gamma^{EELS}(\omega) = \frac{-1}{\pi\hbar\omega}\text{Re}\iiint \vec{E}(\vec{r},\omega)\cdot\vec{J}^*(\vec{r},\omega)d^3r \qquad (3)$$

where $\vec{J}(\vec{r},\omega)$, is the current density function in the frequency domain. Eq. (3) is further simplified to eq. (2), by using the nonrecoil approximation as $\vec{J}(\vec{r},\omega) = \Im\vec{J}(\vec{r},t) = -e\delta(x-x_0)\delta(y-y_0)\exp(i\omega z/v_e)$ [61]. Introducing Eq. (3) has the advantage that the so-called Poynting theorem can be used to link the EELS outcome to that of the photon-generation probability, where the latter is given by integrating the Poynting vector as $\Gamma^{PG}(\omega) = \frac{1}{\pi\hbar\omega}\iint \text{Re}\left\{\vec{\tilde{E}}(\vec{r},\omega)\times\vec{\tilde{H}}^*(\vec{r},\omega)\right\}\cdot\vec{ds}$. The difference between photon-generation probability and loss probability is the so-called absorption spectrum [61]. Note that $\Gamma^{PG}(\omega)$ is directly related to the CL spectra whenever the far-field radiation is considered, for which $\vec{\tilde{H}}(\vec{r},\omega) = (\omega\mu_0)^{-1}\vec{k}_0\times\vec{\tilde{E}}(\vec{r},\omega)$, where $\mu_0$ is the free-space permeability [50]. In this regard, CL is complementary to EELS in studying the near-field of nanophotonic excitations, and can be used to map radiative modes [62, 63]. CL detection systems incorporated in a TEM can serve for EELS versus CL comparison [64]. Advanced CL systems in a scanning electron microscope (SEM) have been recently introduced with the ability to perform CL polarimetry [53].

A great improvement in interpreting the outcomes of low-loss EELS has been achieved by linking EELS to the Green's function and photonic local density of state (PLDOS) [59]. In this context EELS is related to the PLDOS projected along the electron trajectory. In fact, Eq. (2) demonstrates that EELS can be used to directly detect the electric field component projected along the electron trajectory. Pioneered by the work of J. Nelayah et al [65],EELS is applied to many nanophotonics systems to understand near-field distribution and resonant energies of optical modes, especially localized plasmons and plasmon polaritons (see Figure 1) [66-80]. Breathing modes of nanotriangles have emerged as an interesting nonradiative mode, at least when the size of the structure is small enough [67]. Zhu and coworkers probed the PLDOS of a void structure in an Al matrix, and decomposed the full PLDOS into edge and surface plasmons [76]. Using EELS it was demonstrated that Long-range plasmon polaritons propagating in an ultrathin nanowire are robust to the discontinuities induced by bends [72]. Guzzinati et al showed that a properly shaped electron beam can be used to probe symmetrically decomposed modes of plasmonic nanostructures [70]. Acquiring momentum-resolved energy-loss spectra of thin aluminium nanodiscs, Shekhar et al. proved a substantial difference between thin films and nanodics, due to the mode confinement in nanodiscs [73]. The effect of the substrate on the spatial distribution of localized plasmons has been also extensively studied using EELS [68]. Bellido et al. investigated the coupling between adjacent nanoparticles in close proximity to each other, which causes the formation of gap plasmons [66]. Electron beam excitation of exotic toroidal moments in oligomer nanocavities has been also investigated, and proved to contribute negligibly to the far-field radiation in the direction normal to the surface [71]. Formation of wedge hyperbolic polaritons, which are long-range excitations of hyperbolic polaritons propagating along wedges, via coupling between two adjacent edge polaritons, have been also studied in $Bi_2Se_3$ nanoflakes [74]. Finally, considering rotationally symmetric plasmon polaritons of mesoscopic gold tapers, it was shown that phase-matching between the near-field of the electron and excited plasmons can be captured as individual resonances in EELS energy-distance maps [75, 81]. Indeed, it should be recalled that the momentum selection rule $k_z = \omega/V$, as noticed in eq. (2), restricts EELS to map only those photonic states which can afford the necessary momentum. In mesoscopic samples, phase-matching (synchronization) between the electron's self-field and near-field distributions can be manifested in EELS as *new resonances* in structures which support optical modes with higher order angular momentums [75] or in optical gratings [82], (see Figure 1j). Especially for a grating with the period *L*, phase-matching condition will result in $\omega/V_e = k_0 \sin\theta + 2m\pi/L$, where $k_0$ and $\theta$ are respectively the emitted photon wave number and its direction of propagation with respect to the axis of the grating, and $m = 0, 1, 2, \ldots$ is the diffraction order of the grating. This geometrical condition almost precisely describes the criterion for the emission of coherent Smith-Purcell radiation [82-84], which can interfere with electron-induced plasmon radiation [85]. Note that the Smith-Purcell effect is widely accepted as the radiation from the electron beam itself. In the experiment involved, essentially a single electron is expected to arrive at each given time in a TEM; despite this fact however, the Smith-Purcell radiation is a coherent emission process. This concept is also correct for transition radiation [50, 86], which occurs due to the fast annihilation of the created dipole in interaction of a swift electron with its image charge at the surface of a metal.

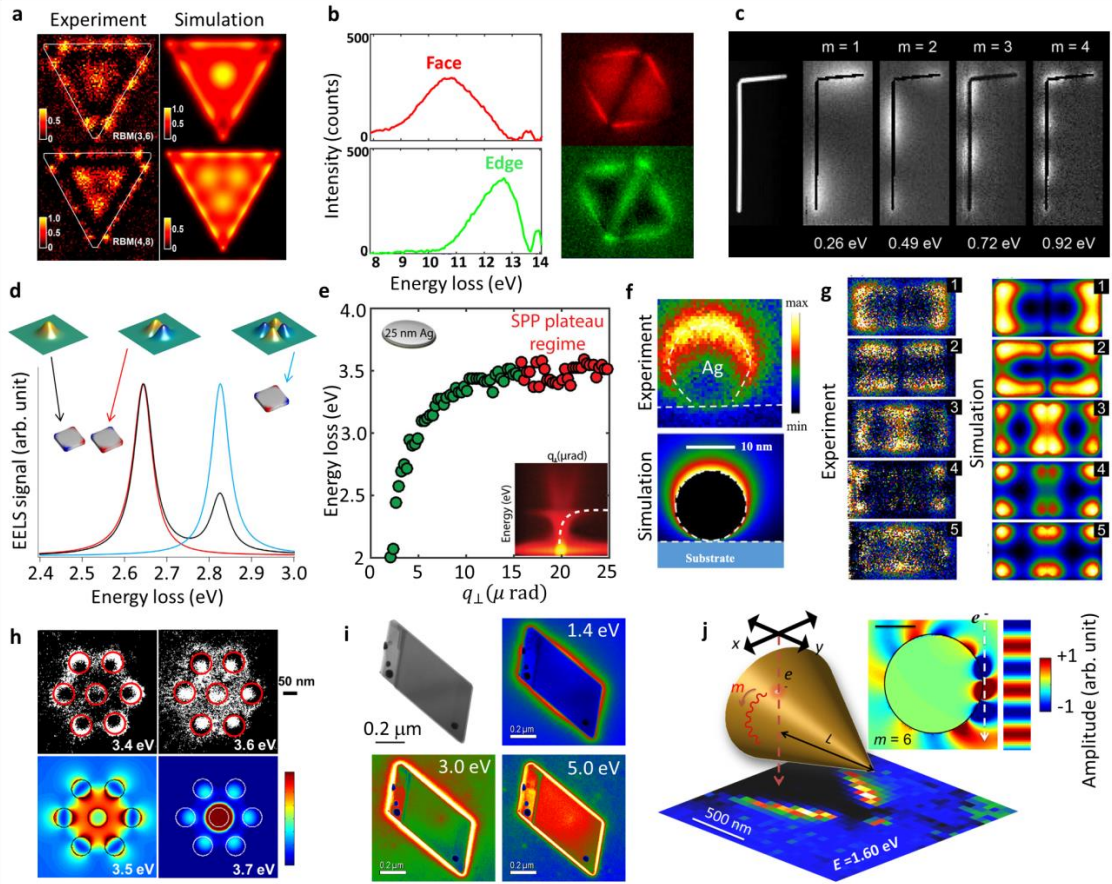

**Figure 1**. EELS investigation of optical excitations in several nanostructures: (a) plasmonic edge and breathing modes (from [67]); (b) edge and face modes of topologically enclosed void structures in an Al matrix (from [76]); (c) long range plasmons in bent silver nanowires (from [72]); (d) dipolar and quadrupolar optical modes of a gold qubic nanostructure decomposed with shaped electron beams (from [70]); (e) plasmonic modes of thin silver discs acquired using momentum-resolved EELS (from [73]); (f) localized plamons of nanospheres in the presence of a substrate (from [68]); (g) coupled edge and gap plasmons in adjacent silver nanoparticles (from [66]); (h) toroidal, azimuthal, and transvers plasmonic modes in voild oligomers (from [71]); (i) wedge hyperbolic polaritons and Dyakonov waves in Bi2Se3 nanoparticles (from [74]); (j) plasmonic modes with higher angular momentum orders in mesoscopic gold tapers (from [75]).

*2.2 Electron recoil*

The applicability of EELS for mapping the photonic local density of states is greatly based on the nonrecoil approximation. It is interesting however to notice that as the electron passes by the neighborhood of nanostructures even in an aloof experiment, the electromagnetic excitations will act on the electron and change its trajectory. Thus, the electron should receive longitudinal as well as transverse recoils because of the Coulomb potential or - in the retarded picture - the Lorentz force. The advantage of the non-recoil approximation is however that the change in momentum (both transverse and longitudinal) during the interaction is insignificant to the final probability and spectral distribution of the interaction, as will be discussed later. Moreover, the fact that better justifies the correctness of the non-recoil approximations for EELS is the weak dependence of the loss spectrum on the incident energy across the entire range of the loss spectrum.

In a combined laser and electron microscope setup, when the structures are illuminated by an external laser light, this recoil can be however better observed. However, to the best of our knowledge, such an experiment has been not yet reported for the electron-induced polarizations in a single structure, in the absence of a laser excitation.

In addition to experiments, numerical studies should be also employed for a better understanding of the interaction of electrons with optical near fields, both in the presence and in the absence of external laser excitations, in particular beyond the nonrecoil approximations. In a classical approach, one might combine the Lorentz and Maxwell equations in a self-consistent way, as routinely employed in particle-in-cell (PIC) numerical procedures [87, 88]. Indeed the success of the PIC method in simulating many physical processes in plasma physics [89, 90], free-electron lasers [91-93], accelerators [92, 93], and in general systems of interacting electromagnetic fields and charged particles has initiated research about self-consistent simulation approaches.

We combined a finite-difference time-domain (FDTD) electromagnetic solver [94, 95] with a Lorentz equation solver, to simulate the interaction of a single relativistic electron with the electron-induced excitations in nanoparticles. Apparently, the system should be treated in a self-consistent way, as the electron first induces the near-field and is then scattered thereof by the self-induced polarization. The electromagnetic field components are assigned to fixed grid points of the FDTD simulation domain (see Figure 2a), whereas the particles are tracked in the continuous domain. The combined equations can be written as

$$\nabla \times \vec{E}(\vec{r},t) = -\frac{\partial \vec{B}(\vec{r},t)}{\partial t}$$

$$\nabla \times \vec{H}(\vec{r},t) = \frac{\partial \vec{D}(\vec{r},t)}{\partial t} + \vec{J}_{el}(\vec{r},t)$$

(4a)

for the electromagnetic fields and

$$\frac{d}{dt}\gamma m \vec{V}_e = \int \rho(\vec{r},t)\left(\vec{E}(\vec{r},t) + \vec{V}_e \times \vec{B}(\vec{r},t)\right)d^3r$$

$$\frac{d\vec{r}_e(t)}{dt} = \vec{V}_e$$

(4b)

where $\vec{E}, \vec{D}, \vec{H},$ and $\vec{B}$ are electric field, magnetic field, displacement vector, and magnetic flux density respectively. $\gamma$ is the Lorentz factor, $m$ is the electron mass, $\vec{r}_e(t)$ is the electron trajectory, and $\rho(\vec{r},t)$ is the electron charge density. The current distribution at the boosted frame is computed as $J'^{\alpha} = \frac{\partial x'^{\alpha}}{\partial x^{\alpha}} J^{\alpha}$, at which $J^{\alpha}$ is the four vector current density distribution in the laboratory frame. In theory, the charge distribution of an electron is perfectly approximated by the Dirac-delta function ($\rho(\vec{r},t) = -e\delta(\vec{r} - \vec{r}_e(t))$), as understood from the Coulomb potential. In practice, however, an electron toy model is often introduced [96], or the charge distribution in the continuous particle space is mapped to the grids of the electromagnetic solver using an extrapolation technique [88]. We used a symmetric Gaussian charge distribution as $-q(2\pi W)^{-3}\exp\left(-0.5|\vec{r} - \vec{r}_e(t)|^2/W^2\right)$ where $W$ is the broadening of the charge distribution. The electromagnetic field in the laboratory frame in free space is given by [97]

$$E'_{z,inc}(\rho',z';t) = \frac{-q\gamma(z'-\beta ct')}{\varepsilon_0\left(\sqrt{2\pi}\right)^3 W\left(\rho'^2+\gamma^2(z'-\beta ct')^2\right)}\exp\left(-\frac{1}{2}\rho'^2+\gamma^2(z'-\beta ct')^2/W^2\right)+$$

$$\frac{-q\gamma(z'-\beta ct')}{4\pi\varepsilon_0\left(\rho'^2+\gamma^2(z'-\beta ct')^2\right)^{\frac{3}{2}}}\mathrm{erf}\left(\left(\rho'^2+\gamma^2(z'-\beta ct')^2\right)^{\frac{1}{2}}/\sqrt{2}W\right)$$

$$E'_{\rho',inc}(\rho',z';t) =$$

$$\frac{-q\gamma\rho'}{\varepsilon_0\left(\sqrt{2\pi}\right)^3 W\left(\rho'^2+\gamma^2(z'-\beta ct')^2\right)}\exp\left(-\frac{1}{2}\frac{\rho'^2+\gamma^2(z'-\beta ct')^2}{W^2}\right) \quad (5)$$

$$+\frac{-q\gamma\rho'}{4\pi\varepsilon_0\left(\rho'^2+\gamma^2(z'-\beta ct')^2\right)^{\frac{3}{2}}}\mathrm{erf}\left(\frac{\left(\rho'^2+\gamma^2(z'-\beta ct')^2\right)^{\frac{1}{2}}}{\sqrt{2}W}\right)$$

where propagation along the *z*-axis has been considered. $\mathrm{erf}(\ )$ is the error function, $(\rho',z')=(x',y',z')$ is boosted coordinate system, and $\vec{\beta}=(0,0,v_e/c)$. The self-interaction, which can be modeled by $\iiint \vec{E}'_{inc}(\vec{r}',t')\cdot\vec{J}'(\vec{r}',t')d^3r'$ is then identically zero, thanks to the symmetry of the field components. In this way, the total field rather than the scattered fields can be used in eq. 4(b). However, numerical implementation of eq. (5) inside an electromagnetic solver introduces additional errors. To cancel the interaction of the electron with its own field, at each time loop we calculate the total field using eq. 4(a) and use $\vec{H}^{sca}=\vec{H}^t-\vec{H}_{inc}$ and $\vec{E}^{sca}=\vec{E}^t-\vec{E}_{inc}$. For an electron interacting with a triangular gold nanoprism with a thickness of 50 nm, the induced electromagnetic field components at a given time are shown in Figure 2b. Modulations of the electron velocity both in the transversal and in longitudinal direction, as well as the total change of the velocity are shown in Figures 2c and 2d respectively.

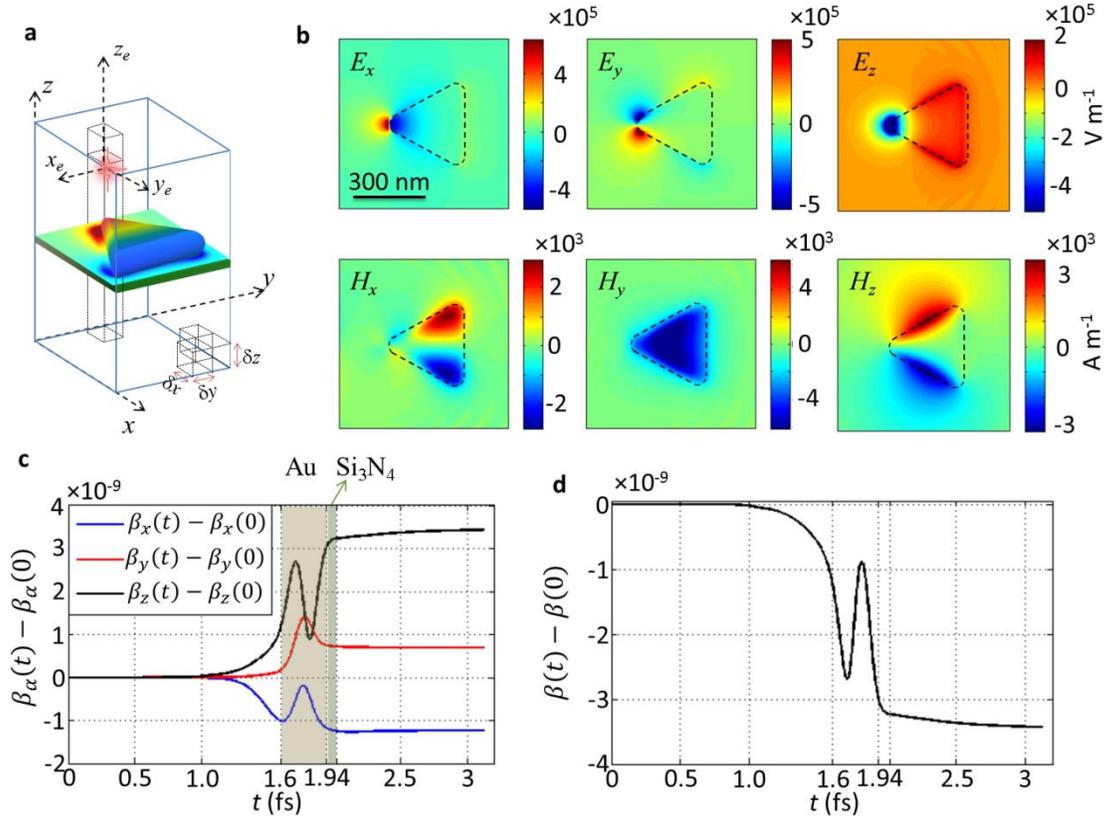

**Figure 2**. Electron at kinetic energy of 200 keV interacting with a single gold nanotriangle prism at 10 nm away from the edge of the structure. (a) Simulation setup including an FDTD electromagnetic field solver and a particle-in-cell tracker. (b) The spatial distribution of induced time-dependent electromagnetic field components at $t = 1.8$ fs, at the plane 5 nm above the gold prism. Modulation of (c) individual and (d) total electron velocity components versus time.

Interestingly, the change of the electron velocity is initiated even before the electron reaches the structure, when it enters the near-field domain. The near-field domain is defined by the region in the vicinity of a given nanostructure at which the momentum of light is larger than the free space momentum. This associated short range decay of optical near field results in an inelastic interaction with the electron. Surprisingly however, the modulation in electron velocity is incredibly low, and the final electron velocity is reduced by an amount of only $\delta v_e = 3.45 \times 10^{-9} c$, where $c$ is the speed of light in vacuum. These findings further confirm the appropriateness of the nonrecoil approximation in treating energy-loss probability and CL spectra. Two points should be however noticed here: 1$^{st}$, The final velocity of the electron and the lateral recoil it experiences, both depend on the impact parameter. In general it is expected that the electron receives less recoil at larger impacts. In other words, in the structure considered above, an impact position at the distance of 10 nm away from the structure results in negligible recoil. 2$^{nd}$, In EELS we detect the probability of the electron to lose certain amount of photon energy. The probability amplitudes indeed depend on the impact parameter, but not the resonant energies. Hence we conclude here that it is the form of the effective interaction potential which affects the resonances, and not the impact parameter. As we previously observed, the probability spectra depend on the frequency-dependent electric-field component. We may relate the change in the longitudinal momentum to the magnetic vector potential as $m_0 \delta \tilde{v}_e(\omega) = q A_z(r_e, \omega)$ and hence to the electric field as $E_z(r_e, \omega) = i\omega_{ph} m_0 \delta \tilde{v}_e(\omega)/q$. Using eq. (2) we may then conclude that is the frequency dependence of the velocity modulation which is related to the EELS spectra and not $\delta v_e = v_e(t \to +\infty) - v_e(t \to 0)$.

Although the experienced electron recoil is apparently negligible, electron energy loss detectors can nowadays precisely determine the electron energy loss probabilities up to the limit of few meVs per electron energy [98, 99]. In systems where the phase-matching (synchronicity) condition between the electron and photons is obtained (see Figure 1j), the overall recoil which an electron experiences can be additively manipulated. This can be achieved by an optical grating as well (see Figure 3a). We consider each element of a grating to be composed of two gold nanowires with a 20 nm gap between them, in order to enhance the interaction and also to maintain a symmetric excitation. The synchronicity condition in this case is exactly analogous to that of the Smith-Purcell effect. The criterion for Smith-Purcell radiation in free-space is given by $\omega/v_e = k_0 \sin\theta + 2m\pi/L$, as described in section 2.1. However, we noticed that it is indeed the power which is transmitted into the substrate which is more pronounced in comparison with the power radiated into other directions. Additionally, a planar waveguide incorporated inside the substrate, such as a thin film of $HfO_2$, can be used to guide the electromagnetic radiation along the symmetry axis of the grating. The Smith-Purcell radiation can be hence coupled into the propagating modes of the waveguide, which are themselves decomposed as usual into the TE and TM waves (see Figure 3b). The time of travel of the electron between two adjacent grating elements is given by $\delta t_e = P/v_e$, while for the emitted photons $\delta t_{ph} = (\beta_r + 2m\pi/P) P/\omega_{ph}$, where $P$ is the period of the grating, $\beta_r$ is the phase constant of the propagating mode of the waveguide, and $m$ is the diffraction order. In order to have constructive interference of the generated photons from the interaction of the electrons with the grating elements, the criterion $\delta t_e = \delta t_{ph}$ should be satisfied. We note however that in contrast with the Smith-Purcell effect in free space, we can even satisfy this criterion with $m = 0$ (see Figure 3b), which greatly enhances the efficiency of the photon generation process. Figure 3c shows the induced *x*-component of the electric field versus time along the electron trajectory. The excited field from the interaction of the electron with the adjacent element is synchronous with the electron velocity and mostly propagating at velocities exceeding the velocity of light in free space. The calculated EELS spectrum sustains a double peak at energies of E=1.9 and 2.1 eV, which correspond to the synchronicity condition ($\beta_r = \omega_{ph}/v_e$) with the lowest order TM and TE modes, respectively (Figures 3d and 3b). As the electron travels at a distance of 5 nm above the grating elements, the overall experienced recoil from the interaction of the electron with the grating is linearly accumulated, and is 3 orders of magnitude higher than the single interaction demonstrated in Figure 2. This is partly due to the incorporation of gap plasmons between adjacent nanorod antennas and partly because of the satisfaction of a synchronous excitation criterion. Comparison between EELS and photon generation spectra however demonstrate that only a small percentage of the photons are coupled to the radiative and propagating photons. In fact, most of the energy is dissipated inside the metallic elements. Nevertheless, the radiation is mostly coupled to the propagating modes of the waveguide, rather than to far-field radiation modes (see Figure 4).

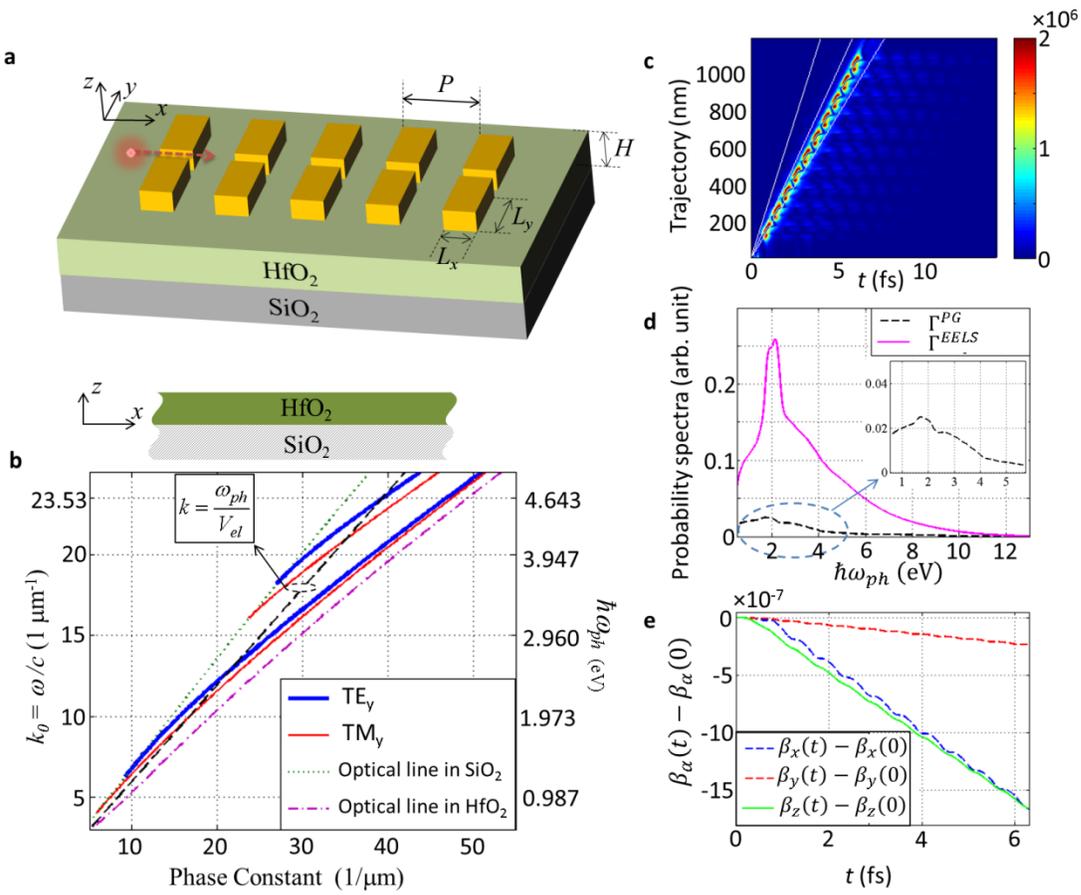

**Figure 3**. Guided Smith Purcell radiation. (a) Interaction of an electron at the velocity of $v_e = 0.598c\,\hat{x}$ with a grating positioned upon a planar waveguide. The grating is composed of gold nanorods with $P$ =185 nm, $L_x$=50 nm, $L_y$ =147 nm, $H$ =180 nm, and the gaps between two nanorods are 20nm. (b) Dispersion of the propagating modes of the waveguide, which can be used to meet the synchronicity condition. (c) Induced x-component of the electric field versus time and electron trajectory. (d) EELS and photon generated probability spectra. (e) Relative velocity of the electron versus time.

The experienced electron recoil can be also controlled using only two nanoantennas coupled to a ridge waveguide [61]. In such a configuration, the localized plasmons excited near the first interaction point is coupled to the plasmon polaritons of a ridge waveguide. A second nanoantenna is positioned along the electron trajectory as shown in Figure 5a, which supports the second interaction. The plasmon polaritons of the first excitation point are then guided towards the second interaction point. They further interfere with the electron induced plasmons of the second nanoantenna either constructively or destructively, depending on the distance $\Delta x$ between the two nanoantennas. The simulated electric field versus time along the electron trajectory for $\Delta x = 327\,\text{nm}$ and $\Delta x = 130\,\text{nm}$ demonstrates this interference phenomenon (compare Figures 5b and 5c). Figures 5d and e show the induced *z*-component of the electric at a given time for $\Delta x = 327\,\text{nm}$ and $\Delta x = 130\,\text{nm}$, respectively. For the former case both nanoantennas resonate out of phase. However they excite the plasmon polaritons of the ridge wave guide in a constructive way. The overall experienced transverse recoil for this case is then doubled, in comparison with the experienced recoil from a single nanoantenna. In contrast, for nanoantennas with a distance of $\Delta x = 130\,\text{nm}$ both nanoantennas resonate in phase. The plasmon polaritons that are excited near the first interaction point, however, reach out of phase with the electron induced plasmons of the second antenna and – thus - interfer destructively. The recoil that the electron receives along the transverse direction (y-direction in this case) is then less than that of the previous case with $\Delta x = 327\,\text{nm}$ (see Figures 5f and g).

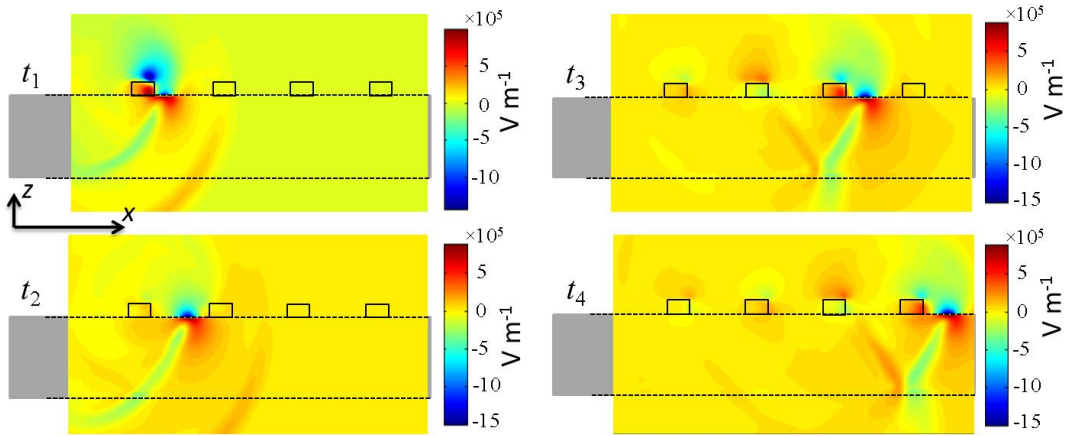

**Figure 4**. Snapshots of the *z*-component of the electric field induced by the interaction of an electron at the velocity of $v_e = 0.598c\,\hat{x}$ with the structure shown in Figure 4a ($t_1 < t_2 < t_3 < t_4$).

### 3. Photon-induced domain

The recoil experienced by the electron can be greatly enhanced both in the transverse and in the longitudinal directions, when an external laser field stimulates the electron photon interactions. The former is well discussed within the elastic Kapitza-Dirac effect [1], whereas the latter is better categorized within the topics of electron energy-gain spectroscopy [100] (EEGS) initiated by Howie, Garcia de Abajo, and Kociak, and PINEM explored by Zewail and coworkers [12]. Recently, the possibility of exploiting plasmon polaritons to manipulate both the transversal and longitudinal recoils of free- electrons has been theoretically explored [101]. The Kapitza-Dirac effect, which is known as the scattering of electron waves by a standing light wave in vacuum, has been long discussed as an interesting example of particle-wave duality of both matter waves and photon waves. This duality is often stated in the literature as the standing-wave light pattern plays the role of a crystalline material lattice, hence leads to the diffraction of electron waves [102]. In the free-space Kapitza-Dirac effect, it is often accepted that it is the ponderomotive potential which is responsible for the scattering of electrons by the vector potential of the light (the Hamiltonian of interaction can be written as $H_1^{int} = \left(e^2/2m_0\right)|\vec{A}|^2$). In the scattering of electrons by plasmons however, the interaction is most often taken as $H_2^{int} = \left(-i\hbar e/m_0\right)\vec{A}\cdot\vec{\nabla}$ [103]. $H_2^{int}$ is indeed an indication of single photon emission and absorption, which is not possible in free-space because of the violation of the energy-momentum conservation [102]. In the vicinity of nanostructures, single-photon processes can happen and are indeed more probable than two-photon processes indicated by $H_1^{int}$. However, this is only true to the first order within the Born approximation scattering theory and at certain field amplitudes, $H_1^{int}$ and $H_2^{int}$ both contribute significantly to the overall Hamiltonian, and should be considered. By approximating $\vec{E} = i\omega_{ph}\vec{A}$ and $\vec{\nabla} \to i\vec{k}_e$, where $k_e$ is the electron wave number, and $\omega_{ph}$ is the angular frequency of the light, the critical electric field amplitude is obtained as $E_c = \left(2m_0\omega_{ph}/e\right)V_e$. Apparently, for having $H_1^{int}$ as the dominant part, $E_0 \gg E_c$. For relativistic electrons and photons at the visible range, this condition demands very large amplitudes for the laser field. For slow electrons this critical value is however more easily achieved. Nevertheless, for relativistic electrons and light at moderate intensities and at optical frequencies, $H_2^{int}$ is the dominant term [12].

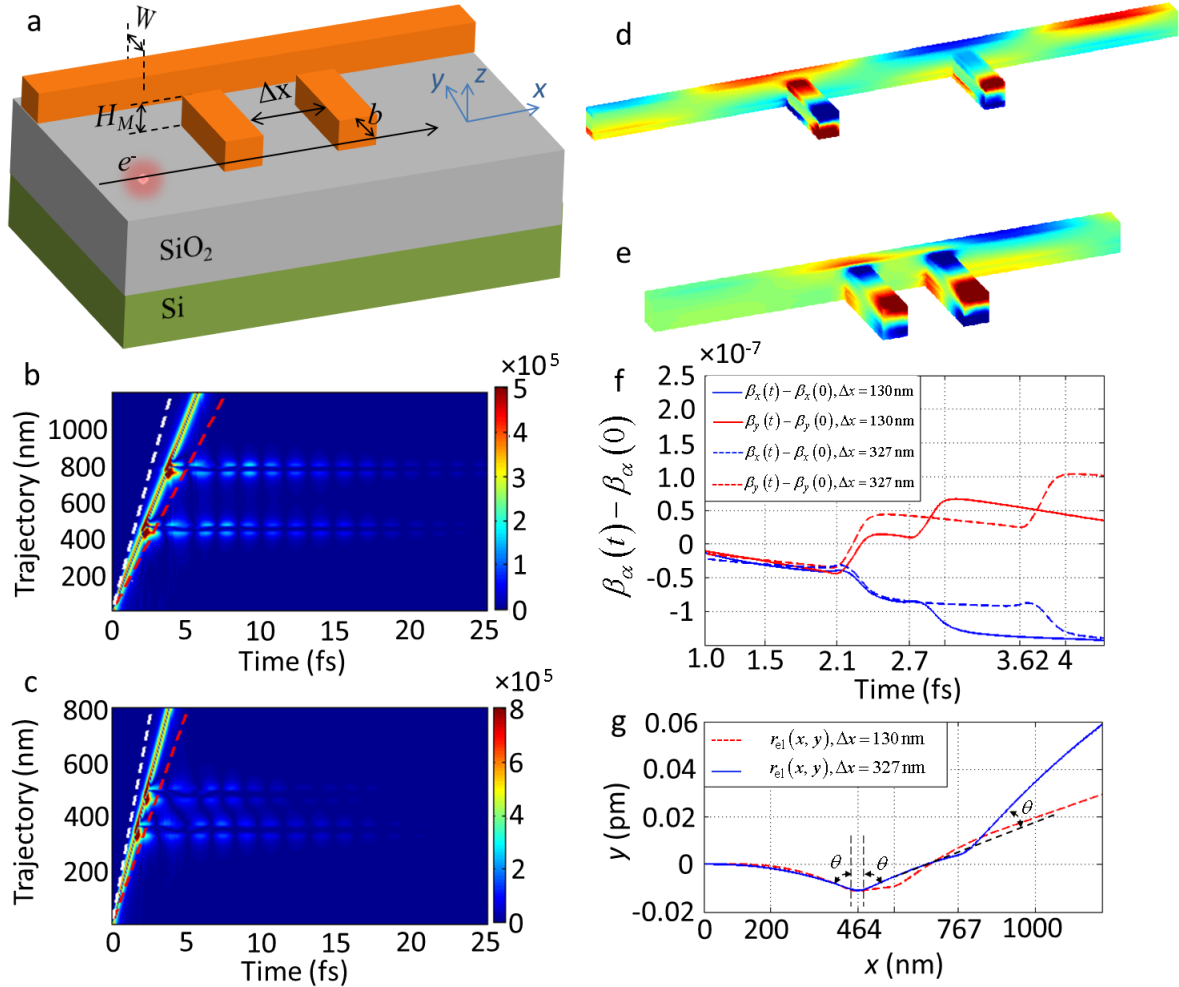

**Figure 5.** Controlling the electron recoil with nanoantennas [61]. (a) Topology of the structure composed of gold nanoantennas coupled to a ridge waveguide interacting with an electron at the kinetic energy of 200 keV propagating along x-axis at 2 nm away from the nanoantennas. $W = 30$ nm, $H_M = 30$ nm, and the length of the nanoanatennas is 147 nm. Induced electric field along the electron trajectory for (b) $\Delta x = 327$ nm and (c) $\Delta x = 130$ nm. z-component of the induced electric field for (d) $\Delta x = 327$ nm and (e) $\Delta x = 130$ nm. (f) The relative velocity of the electron ($\beta = v_e/c$) along the x- and y directions. $\theta = 1.57 \times 10^{-8}$ rad is the scattering angle of an electron in interaction with a single nanoantenna.

### *3.1 Electron energy – gain spectroscopy*

In 2008, Garcia de Abajo and Kociak introduced EEGS as a tool to improve the energy resolution of electron microscopes [100]. In an EELS setup, electrons undergo spontaneous emission in interaction with nanostructures and lose energy. Hence in the presence of external continuous – wave (CW) laser excitations, stimulated photon emission and photon absorption processes will be also possible. In other words, at sufficiently high laser energies, electron energy-gain and energy-loss processes are both possible. Based on Fermi's golden rule, they obtained the energy gain expression for single photon-electron interactions as

$$\Gamma^{\text{EEGS}}(\omega) = \left(\frac{e}{\hbar\omega}\right)^2 \left|\tilde{E}_z(x_0, y_0, k_z = \omega/v_e; \omega)\right|^2 \qquad (6)$$

which is related to the intensity of the electric field, in contrast with the EELS formalism (see Eq. (2)). Obviously, using the classical approach, as used for obtaining the EELS probability, an incorrect relation for the gain probability is derived, for which $\Gamma^{\text{EEGS}}$ is linearly related to the electric field amplitude [100]. Moreover, negative probabilities can be obtained, if eq. (2) is incorrectly used for calculating the energy gain/loss spectra in the presence of an external laser excitation.

## 3.2    *Photon-induced near-field electron microscopy*

In PINEM, a pulsed laser excitation is synchronized by the electrons emitted from a photoemission electron gun, within a photon-pump and electron-probe time-resolved spectroscopy apparatus (see Figure 6a). Observations of energy gain and loss processes in a series of spectacular experiments carried out in the group of Zewail and co-workers, demonstrated a series of loss and gain peaks in the spectra of carbon nanotubes (see Figure 6b) [9, 12, 104]. Indeed, those experiments demonstrated the possibility of multiple-photon gain and loss processes in addition to the single-photon processes. Similar multiphoton processes were observed in the interaction of electrons with protein vesicle (see Figure 6c). To describe the effects theoretically, a scattering theory based on the semiclassical propagators for Schrödinger equation has been developed by Park and coworkers, and demonstrated to be in a good agreement with experiments [12]. Followed by the pioneering experiments mentioned above, many other groups have already advanced the PINEM technique towards realizing better temporal and spatial resolutions. We outline here only works based on spectroscopy; for ultrafast electron diffraction the readers are referred to the review by Zewail [105] and more recent works [106]. Feist et al. have recently demonstrated that the strong resonant interaction between the laser beam and electrons in the vicinity of a nanostructure can be used to induce multilevel Rabi-oscillations, with equal Rabi-frequencies between the levels, into the electron wave function (see Figure 7a) [17]. The probability for the transition between the photonic levels in an equal-Rabi case is given by Bessel functions, in contrast with the Rabi oscillation in a harmonic oscillator system which is describable using Poisson distribution [107]. This behavior results in a strong oscillation in the intensity of the observed gain and loss peaks, rather than a monotonous decrease in the intensity versus the order of electron-photon interaction processes. They described their observation using a quantum optical framework on annihilation and creation ladder operators for the plasmon modes which commute (neglecting the spontaneous emission). Based on this approach, they proposed a more accurate description of the kinematic and dynamical processes mentioned by Zewail in a previous work (Figure 6 c). This approach in general, which is called boson sampling, holds the promise of solving intractable problems which cannot be efficiently simulated using classical computers [108, 109]. Additionally, they proposed a quantum coherent manipulation of free-electron states using optical metrologies, which were further experimentally confirmed and expanded by demonstrating a Ramsey-type phase-control of free-electron motions [28, 110].

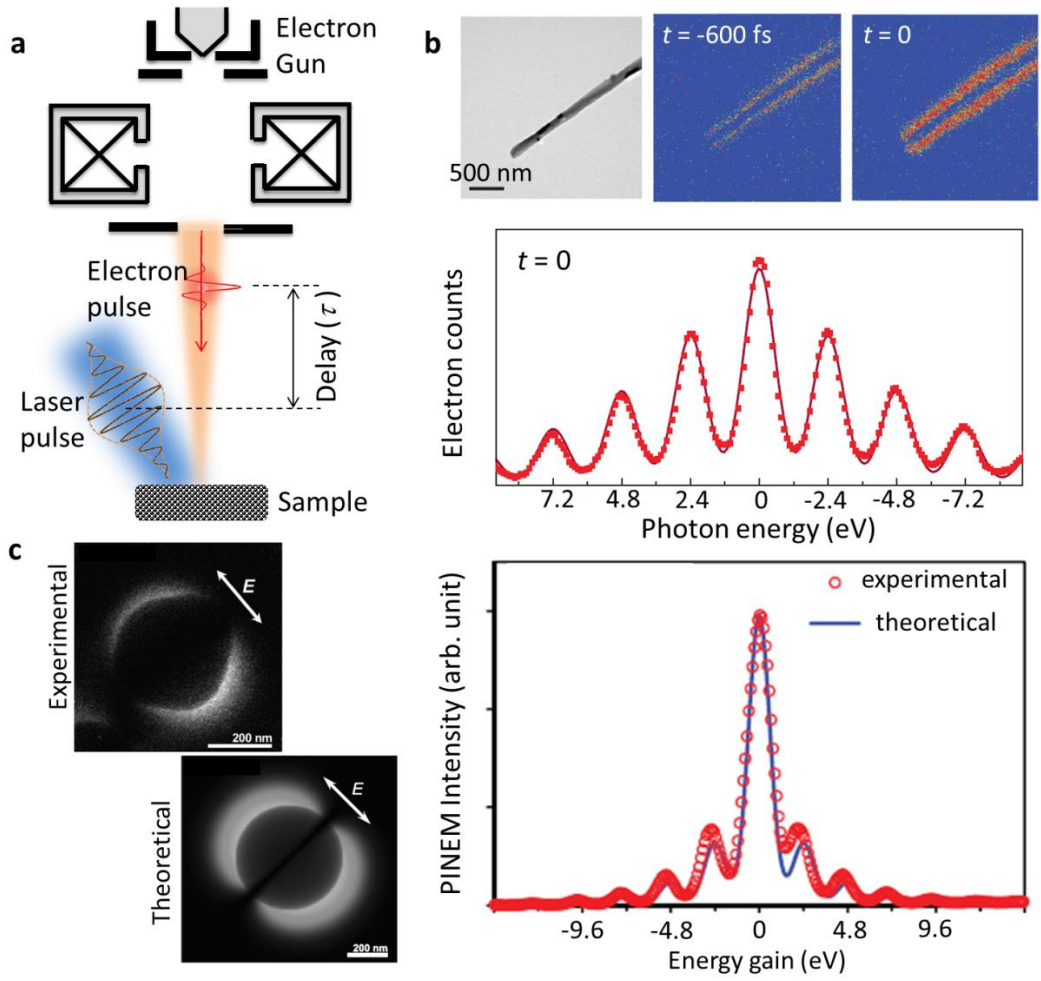

**Figure 6.** Photon-induced near-field electron microscopy. (a) The setup is a pump-probe spectroscopy apparatus were the pump is a laser pulse, and the probe is an electron pulse. The delay between these two pulses is precisely controlled using an optical delay line. (b) top: PINEM images on a single carbon nanotube, shown at two different delays, bottom: PINEM intensity versus energy at delay time $t = 0$ [9]. (c) Experimental and theoretical PINEM time-resolved images (left), and spectra (right) of a single protein vesicle with a radius of 150 nm [12].

Interestingly, such near-field interactions with electron wave packets leads to the generation of a train of attosecond electron pulses propagating in space-time [17], which might be further utilized for ultrafast diffraction [111]. Piazza et al. have recently demonstrated up to 9 orders of photon emission and absorption peaks in a PINEM time-energy map of silver nanowires (Figure 7b) [18]. They also highlighted an interesting aspect of PINEM for recording the spatial dependence of individual loss and gain peaks [112]. Finally, Ryabov and Baum have shown that indeed transversal electron recoil can be acquired and utilized to map the spatiotemporal distribution of transversal field components, in addition to the longitudinal gain and loss peaks accumulated in the longitudinal momentum of the free-electron waves [113]. They used THz laser pulses and electron pulses as short as 80 fs to respectively initiate and probe the electromagnetic excitations in a split ring resonator (Figure 7c). Their unique observations in this field is intriguing to the field of ultrafast science and electron microscopy in principle, as they clearly demonstrate the advantage of recording the transversal momentum of the electron in addition to its longitudinal momentum in a PINEM setup.

### 3.3  *Ultrafast point projection electron microscopy*

Complementary to the developments of ultrafast techniques in TEMs, ultrafast PPM has been recently appeared as a probe of ultrafast electron dynamics in samples.  PPM holds the advantage of a less complicated setup without the need for massive magnetic lenses in TEMs, though the spatial resolution has yet to be improved to compete with ultrafast TEMs. The key element behind the development of a PPM is the recent progress in controlling the ultrafast dynamics of photoelectrons emitted from sharp tips, either employing adiabatic nanofocusing [38-40] for transferring the grating-coupled light from the shaft at several micrometres away from the apex to the apex itself (see Figure 8a and b), or by directly illuminating the tip apex [21, 114, 115] (see Figures 8c and 8d). The former setup has advantages in bringing the sample closer to the tip, which by itself helps for shorter electron pulses, and achieving higher magnification (see Figure 8b). Using adiabatic nanofocusing, realization of electron pulses as short as 8 fs has been reported [39]. However, as noticed by Mueller et al, dispersion may cause few femtosecond broadening of the emitted electron pulse (see Figure 8c, which shows the interferometric autocorrelation of photoelectrons from the tip) [39]. A certain advantage of PPM is the ability to perform inline electron holography. Electron holography was indeed suggested by Gabor in 1948 [116] as a new technique to overcome the problems of lens aberration and to improve electron microscopy techniques. Still holography in PPM is used to improve the spatial resolution by an order of magnitude [39], thanks to the transversal interference patterns which are observed in the PPM images. Moreover, the comparison between the interference fringes for the case of field-driven and laser-driven tungsten nanotips, demonstrates that the effective size of the source for both cases is not very much different. In other words, the promise of utilizing point sources in PPM holds true for a laser-driven source as well as for a field-emission source (see Figure 8f) [21].

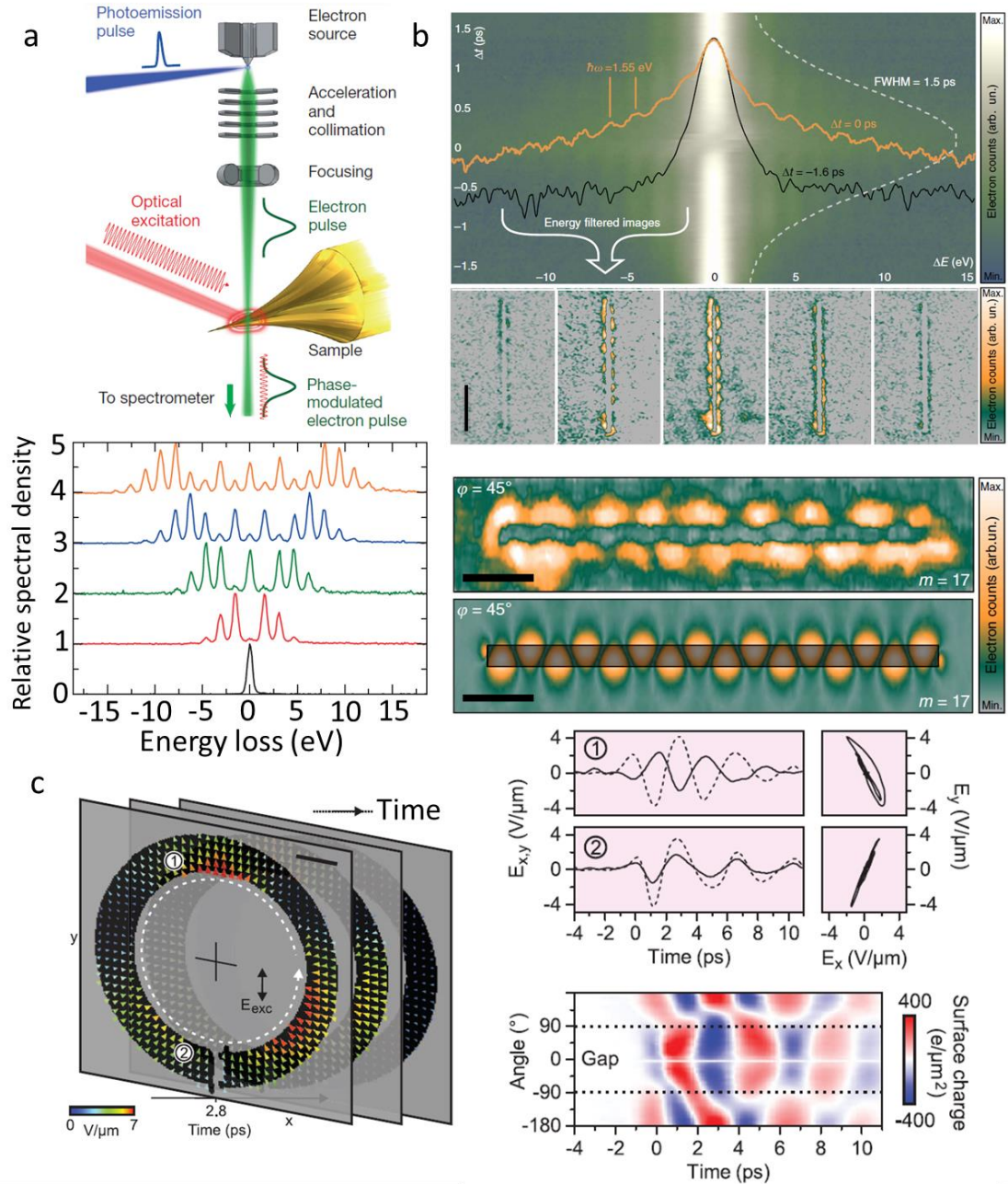

**Figure 7.** PINEM activities in several groups. (a) Top: By precisely controlling the phase and polarization of the incident light, the PINEM setup can be used to coherently manipulate the free-electron states. Bottom: Electron energy spectra for several field amplitudes of 0, 0.023, 0.040, 0.053, and 0.068 V nm-1 [17]. (b) Top: Time-energy phase-space PINEM map of silver nanowires. Bottom: Experimental and numerical PINEM images of the field distribution on an isolated nanowire with light excitation polarized at the angle of 45◦ with respect to its longitudinal axis [18]. (c) Left: Vector distribution of the lateral components of the electric field for a split-ring resonator excited with a THz laser pulse, and probed with electron pulses with only 80 fs duration. Right, top: temporal distribution of the excitations at two distinct test points marked with A and B in the left panel, and also the polarization state of the electric field. Right, bottom: The temporal distribution of the electric field along the inner circumference shown by dashed line in the left panel [113].

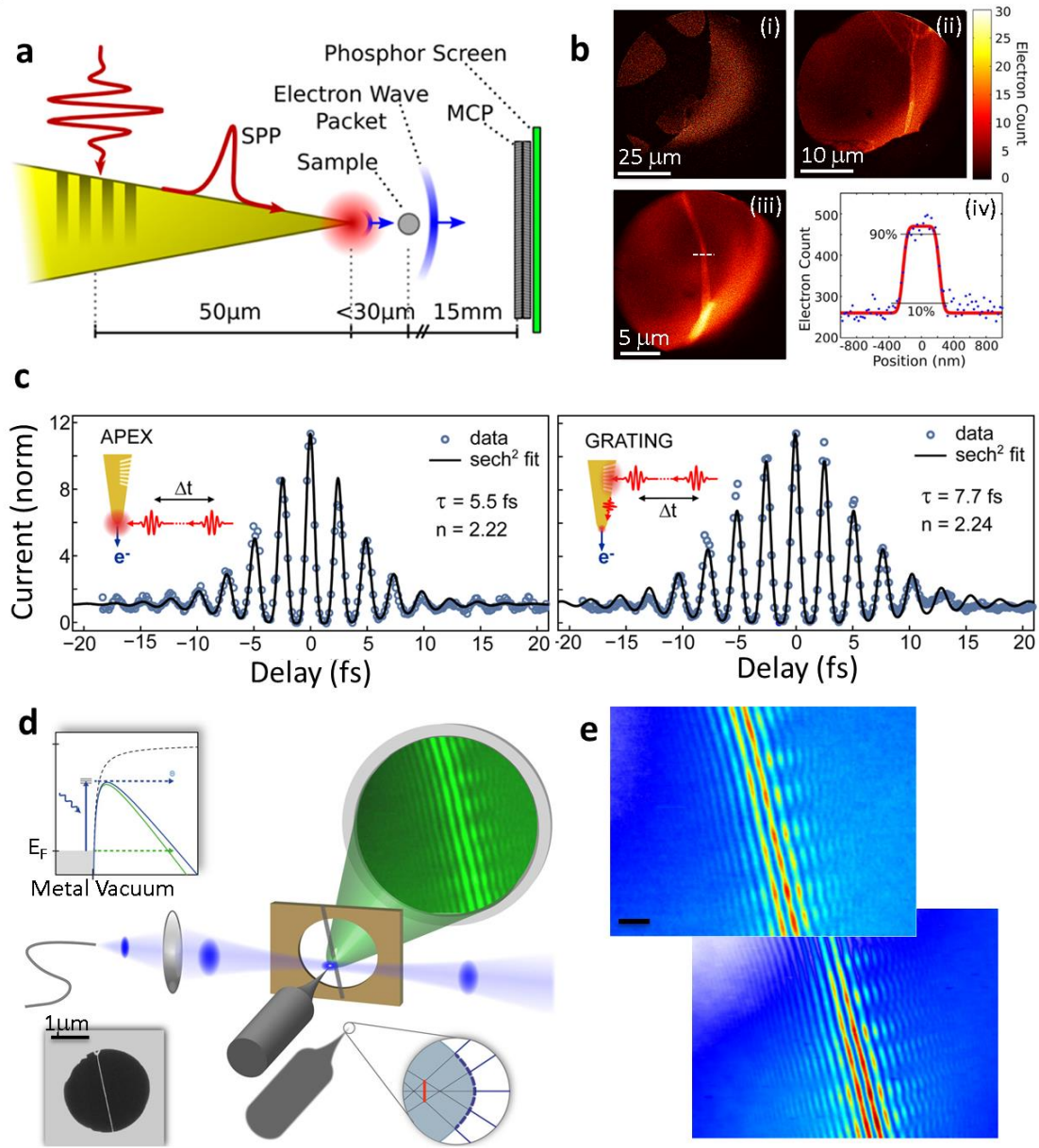

**Figure 8.** PPM with laser driven nanotips. (a) By adiabatically nanofocusing the plasmons from the grating to the apex, the sample can be positioned arbitrary close to the tip, which helps for achieving higher magnifications [40]. (b) (i) to (iii) PPM images from a Ag nanowire recorded by nanofocusing for three different magnifications, and (iv) Cross section along the white line in left lower panel [40]. (c) Comparison of the interferometric autocorrelation of photoelectrons emitted from the sample using either adiabatic nanofocusing or a direct illumination of the tip apex, as shown in the insets [39]. (d) Recording an interference pattern of the electron beams

exciting a carbon nanotube, using PPM [21]. (f) Several interference fringes are observed, both for a laser-driven gun (upper panel), and field-emission gun (lower panel) [21].

*3.4    Time and energy resolutions in ultrafast electron microscopy*

Improving the time resolution of ultrafast techniques in general, and ultrafast electron microscopy in particular, is highly demanded for unravelling charge-transfer dynamics, time evolution of the electromagnetic oscillations in nanostructures, and understanding the behaviour of correlated materials [29]. It has been however increasingly noticed that reaching attosecond time resolution demands the utilization of sophisticated techniques such as highly precise synchronization with microwave cavities [23, 30] or THz cavities [31], to chirp the electron pulse distribution in time-energy phase space. Although electron pulses as short as few fs have been realized, the increase in time resolution is concomitant with the broadening of the pulse in the energy domain which covers a large bandwidth and is beneficial for spectroscopy and investigation of resonant phenomena as well. In addition to the synchronization cavities, inclusion of an appropriate low-emittance field-emitter electron gun utilizing nanolocalized photoemission from a zirconium oxide covered (100) – oriented single crystalline tungsten tip has been also reported [117], which significantly improves the beam specifications. The ultrafast electron microscope facilitated with this tip has been shown to sustain the spatial resolution of 0.9 Å, energy broadening of 0.6 eV, and temporal resolution of 200 fs.

**4.    Photon – assisted domain and spectral interferometry**

Considering electron beams and laser excitations as two individual probes of the sample excitation, two different domains have been already discussed, namely the electron-induced domain and the photon-induced domain. Members of the former domains are EELS and CL techniques, whereas in the latter domain PINEM and ultrafast PPM are obvious examples.

Within the electron induced domain, the electron interacts with the optical modes of the sample which are initially at the ground state. In the classical picture, one can assume that the evanescent field of the electron probe polarizes the sample, and the electron is hence scattered thereof by its induced field (see section 2.1). Interestingly, multiple photon emission processes might also occur and results in multiple loss peaks. This effect has been observed in experiments carried by Powel and Swan as early as 1959 [118, 119], and are also observed in simulations considering retardation effect [120]. Within the photon-induced domain however, the sample is pumped with a strong laser beam into a superposition of number states. In addition to the loss channels caused by the spontaneous emission, stimulated emission and absorption processes will happen as well and indeed in most experimental cases become the only relevant loss and gain channels (see section 3.2). In other words the electron-induced polarization in the sample can be neglected. One might try to classify these two apparently distinct electron-induced and photon-induced domains versus the intensity of the incident laser beam, where at very low laser intensities, still the electron-induced polarization can be probed using EELS. In contrast, at sufficiently high laser intensities, only PINEM excitations are probed by the electron, as the electron-induced spontaneous emission can be neglected. At an intermediate intensity, coherent electron induced radiations should be able to interfere with the laser-induced fields [121] (see Figure 9a). This interference phenomenon then can be probed using EELS (see Figure 9b), and is highly beneficial for recording the spectral phase as an example. The laser field amplitude at which this transition occurs depends on the kinetic energy of the electron beam and its impact parameter. For the plasmonic nanostructure considered here, and for an electron at the kinetic energy of 200 keV traversing the nearfield at 5 nm away from the nanostructure, this field amplitude is approximately $10^5 \sim 10^6$ V/m, as unraveled by the simulation. Interestingly, conventional EELS at the absence of laser excitation is a more useful spectroscopy tool with the ability to probe all the optical modes that the structure sustains, including here the resonances at $E = 0.25$, 1.0, and 1.55 eV. The laser pulse will indeed peak up only the mode at $E = 1$ eV. However, the introduced laser-electron pump-probe approach let us to determine the temporal evolution of selected plasmon excitations.

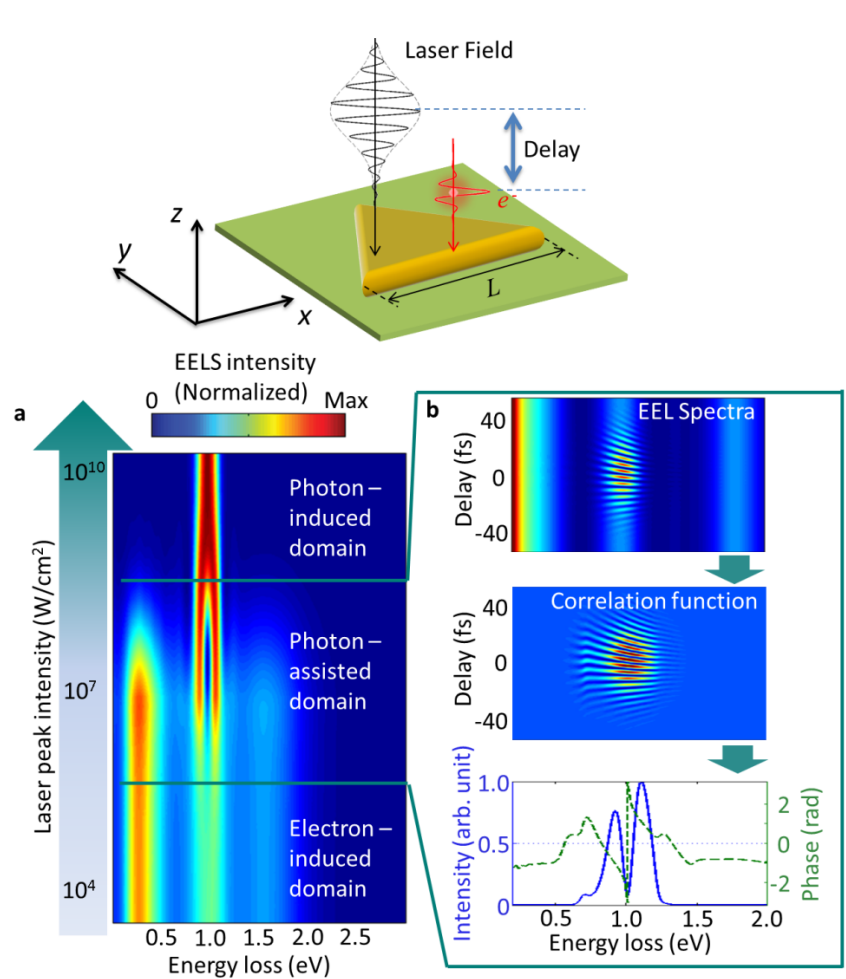

**Figure 9.** Distinguished domains in the interaction of single-electron pulses and laser pulses with nanostructures. An electron at the kinetic energy of 200 keV interacts with a triangular gold nanostructure with the edge length of L = 400nm and thickness of 30 nm, which is positioned upon a $Si_3N_4$ substrate [121]. The structure is pumped by a y-polarized laser pulse with the temporal broadening of 10 fs and the carrier photon energy of 0.98 eV. The electron pulse has a temporal broadening of only 48 as. (a) Electron-induced, photon-assisted, and photon-induced domains are categorized versus the intensity of the laser pulse. (b) within the photon-assisted domain, the strength of the induced laser illumination is at the level of the electron-induced polarization, and can further interfere with it. Additionally, interference patterns can be recorded using EELS and be utilized for recovering the spectral phase.

In order to record such an interference map, ultrashort sub-fs electron pulses are required, which are perfectly synchronized with the laser field [122]. It is due to the fact that the duration of the electron pulses should be shorter than a cycle of the laser excitation, to be able to appropriately probe the gain and loss oscillations along the delay axis [121]. Additionally, only coherent radiations like transition radiation and coherent Bremsstrahlung can interfere with the incident photon pulses. Incoherent CL in interaction of electrons with dielectrics, semiconductors and semimetals, will however hamper the visibility of fringes.

Based on the phenomenon noted above, we have proposed a system for improving the synchronization of the electron and photon pulses upon their arrival at the sample. Instead of triggering the electron pulses with photon pulses using photoemission processes, an inverse approach has been outlined which is based on coherent electron induced radiations like transition radiation [35]. As discussed previously in section 2.2, two nanoantennas coupled to a waveguide can be used to manipulate the experienced recoil by the electron. Depending on the distance between the nanoantennas, the experienced recoil might be enhanced or hampered. We will generalize this concept to the interaction between two dissimilar structures, and replace the plasmonic rib waveguide by free space propagation. In other words, we replace the first nanoantenna with a structure which in interaction with the electron creates coherent radiation which is focused on the sample (see Figure 10a). This structure is called an electron-driven photon source (EDPHS). An example of an EDPHS structure is an inverted superlens (see Figure 10b). This structure offers several degrees of freedom for triggering the electron induced emission towards the realization of a focused and directional radiation; attained by combining the photonic crystal structure with a multilayered hyperbolic material [123]. The radiation from this EDPHS structure is in the form of a transverse magnetic and ultrashort one cycle pulse [35] (see Figure 10b). The second nanoantenna is then replaced by the sample, which the latter is positioned in the focal point of the EDPHS. By changing the distance $L$ between the EDPHS and the sample, the arrival time of the electrons and EDPHS on the sample is controlled. In other words $L$ is correlated with the temporal delay between electron and photon as $\tau = L\left(v_e^{-1} - c^{-1}\right)$, where $c$ is the speed of light in free space. By changing the distance the recoil experienced by the electron and its velocity can be manipulated as shown in Figure 10c. This strong modulation of the velocity hints at the excitation of an ultrashort electromagnetic radiation from EDPHS which can further interact with the electron at a certain distance from the EDPHS. It is not only the electron velocity which can be controlled using EDPHS, but also the EELS and CL spectra are altered by the distance $L$ (see Figure 11a for CL map versus photon energy and $L$). The change in CL spectrum versus $L$ is more pronounced than the changes in the EELS map, which is due to the constructive and destructive interferences of the photons generated from EDPHS and the sample at the detector plane. By finely tuning the distance $L$, an overall interference pattern is recorded which can be used to extract the correlation function (see Figure 11b) and to further extract the spectral phase (see Figure 11c). We define here the time-frequency correlation function as $C(\tau,\omega) = -1 + \Gamma^{CL}(\tau,\omega)/\Gamma^{CL}_{EDPHS}(\tau,\omega)$, where $\Gamma^{CL}(\tau,\omega)$ is the total energy-time (energy-distance) CL map shown in Figure 11a, and $\Gamma^{CL}_{EDPHS}(\tau,\omega)$ is the EDPHS CL spectrum which is acquired as a pre-step and is taken as the reference. $C(\tau,\omega)$ is further converted to a two-frequency correlation function using Fourier transformation, where the latter is employed to retrieve both the intensity and phase of the *electric field* spectrum [35]. It is to be noted that this system can be viewed as an inline holography apparatus, whereas the *images* are recorded in the distance-energy space rather than in the two-dimensional distance-distance space [124, 125]. The retrieved spectrum is the spectra of the electron induced electric field from the sample, relative to the phase of the EDPHS radiation. The plasmon peak excited at 2.9 eV is associated with the breathing mode of the silver disc [126], whereas the peak at 3.6 eV is due to the excitation of bulk plasmons.

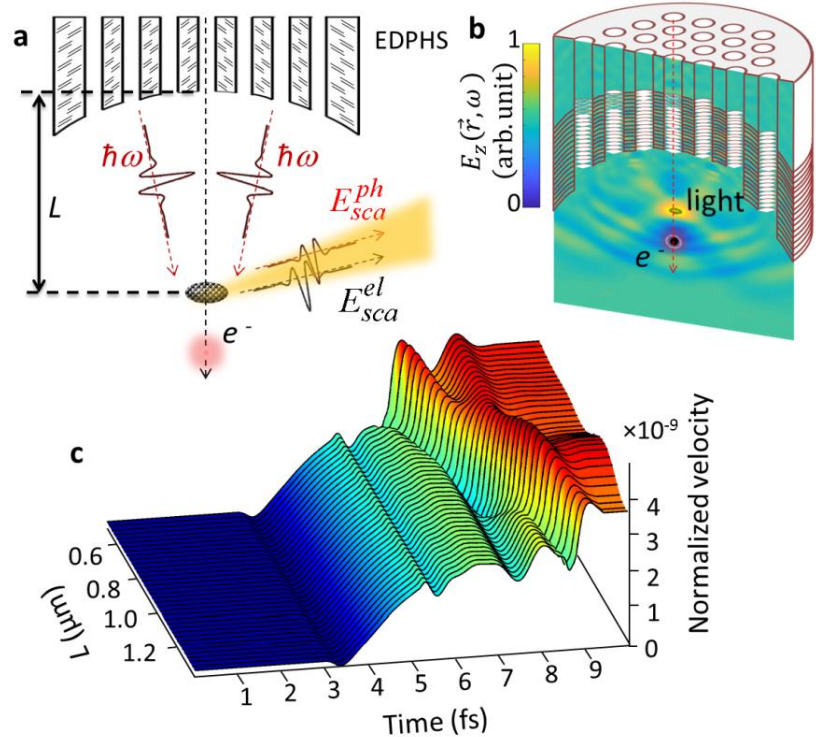

**Figure 10**. Basis of spectral interferometry with electron microscopes [35]. (a) An EDPHS interacts with an impinging electron and emits coherent radiation which is focused onto the sample. The electron-induced field that is scattered from the sample $E_{sca}^{el}$ will interfere with the radiation generated by the EDPHS, $E_{sca}^{ph}$. (b) Design of the EDPHS, comprising an inverted superlens composed of a multilayer structure on the glass and an incorporated void hexagonal photonic crystal. (c) Normalized velocity of the electron versus time of flight of the electron and the distance $L$ between EDPHS and sample.

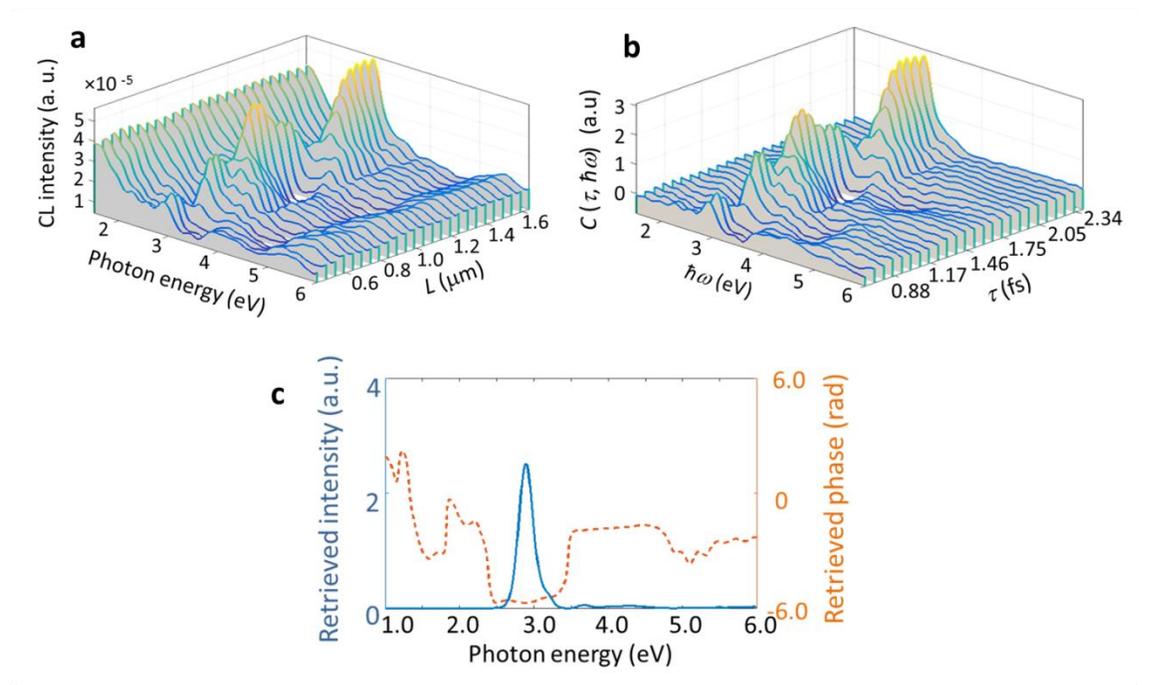

**Figure 11**. Recovering the spectral phase using the spectral interferometry technique [35]. (a) Simulated CL spectra resulting from the interaction of an electron at energy of 200keV with the EDPHS and a thin silver disc (with the thickness of 40 nm) as the sample. The dependence of the surface and bulk plasmon resonances at 2.9 eV and 3.6 eV on the distance $L$ between EDPHS and sample, clearly demonstrates the interference patterns. Extracted (b) correlation function (see text) and (c) spectral phase of the sample excitations versus the phase of the EDPHS radiation [35].

## 5. Simulations beyond adiabatic assumptions

The majority of strong-field effects in light-matter interaction is qualitatively understood using adiabatic approximations [127]. These approximations most often treat the electron gas being unbound to the ions, where the evolution of their wave function in interaction with the laser field is represented by Wolkow states [128]. This approach treats the change in the electron wave function via its propagation in the laser field with a special case of the eikonal approximation where the change of the amplitude is neglected. The Wolkow wave function is given by

$$\psi(\vec{r},t) = \frac{1}{(2\pi)^{\frac{3}{2}}} \exp(i\vec{k}_e \cdot \vec{r} - i\omega_e t)$$
$$\times \exp\left(-i\frac{e}{\hbar m_0} \int_{-\infty}^{t} \left[\hbar \vec{k}_e \cdot \vec{A}(\tau) + \frac{e}{2}|\vec{A}(\tau)|^2 - m_0 \varphi(\tau)\right] d\tau\right) \tag{7}$$

where $m_0$ is the electron mass, $\vec{k}_e = m_0 \vec{v}_e / \hbar$ is the electron wave vector, $\omega_e = \hbar k_e^2 / 2m_0$, and $\vec{A}(t)$ is the vector potential, which is considered to be only dependent on time, as the wavelength of photons are assumed to be much larger than the extent of the whole wave function. The first term in the integrand underpins the acceleration and deceleration of the electron in interaction with light, within the dipole approximation. The second term describes the ponderomotive force on the electron, and the last term is correlated with the change in the phase of the electron via its propagation in the effective electrostatic potential of the light.

### 5.1 Slowly varying approximation

In general however, the time dependent Schrödinger equation including electromagnetic interactions is given by

$$-\frac{\hbar^2}{2m_0}\nabla^2\psi + \frac{-i\hbar e}{m_0}\vec{A}\cdot\vec{\nabla}\psi + \left(\frac{e^2}{2m_0}\left|\vec{A}\right|^2 - e\varphi\right)\psi = i\hbar\frac{\partial\psi}{\partial t} \quad (8)$$

where the Coulomb gauge has been implied, and $\varphi$ is the scalar potential. Note that despite the fact the Schrödinger equation is nonrelativistic, still relativistic corrections can be applied, to be able to employ eq. (8) for understanding the dynamics of relativistic electrons. This is achieved by simplifying the Dirac equation into a scalar form, by ignoring the electron spin [129], which the latter is most often unimportant in electron microscopy. Moreover, Dirac equations introduce formidable complications in electron-beam physics such as Zitterbewegung behavior [130] and spontaneous electron-positron pair creation.

A common practice in electron microscopy, especially when high energy electron beams are involved, is to write the wave function as

$$\psi(\vec{r},t) = \psi_0(\vec{r},t)\exp\left(i\vec{k}_e\cdot\vec{r} - i\omega_e t\right) \quad (9)$$

where $\psi_0(\vec{r},t)$ is the slowly-varying amplitude. Using this approximation, we recast eq. (8) into a more practical formula as

$$\begin{aligned}
&-\frac{\hbar^2}{2m_0}\nabla^2\psi_0(\vec{r},t) - \frac{i\hbar^2}{m_0}\vec{k}_e\cdot\vec{\nabla}\psi_0(\vec{r},t) \\
&+ \frac{-i\hbar e}{m_0}\vec{A}\cdot\left[\vec{\nabla}\psi_0(\vec{r},t) + i\vec{k}_e\psi_0(\vec{r},t)\right] \\
&+ \left(\frac{e^2}{2m_0}\left|\vec{A}\right|^2 - e\varphi\right)\psi_0(\vec{r},t) \\
&= i\hbar\frac{\partial\psi_0(\vec{r},t)}{\partial t}
\end{aligned} \quad (10)$$

By assuming $\vec{\nabla}\psi_0 \ll i\vec{k}_e\psi_0$ (slowly varying approximation (SVA)), eq. (10) is further simplified into the form

$$-\frac{\hbar^2}{2m_0}\nabla^2\psi_0(\vec{r},t) - e\vec{A}\cdot\vec{V}_e\psi_0(\vec{r},t)$$
$$+\left(\frac{e^2}{2m_0}|\vec{A}|^2 - e\varphi\right)\psi_0(\vec{r},t) \qquad (11)$$
$$= i\hbar\frac{\partial\psi_0(\vec{r},t)}{\partial t}$$

Where, in the strong-field approximation, the second term in eq. (10) can be neglected (Strong-field approximation). In order to derive the Wolkow wave functions from eq. (11), we assume that $\psi_0(\vec{r},t) = \psi_0(t)$, which by insertion into eq. (11) and appropriate normalization, leads to eq. (7).

It is already noticed that in the case of energetic and relativistic electron beams, Wolkow states and SVA provide us with the analytical treatment of many experimental observations, from PINEM [12, 131] to electron holography [132]. However, certainly Wolkow wave functions will not provide us with enough insight into more advanced methodologies based on the shaping of the electron wave functions and point projection microscopy with slow electrons. This is due to the reason that Wolkow states are outcomes of an adiabatic approximation where the amplitude of the electron wave function is neglected. Moreover, implicit to eq. (11) is the dipole approximation, which might break down in some realistic situations.

We here propose a test analysis for comparing the solutions to eq. (8) and eq. (11) for the case of the Kapitza-Dirac effect. Our system involves an electron at the velocity of only $v_e = 0.02c$, interacting with two CW Gaussian optical waves at a wavelength of 30nm propagating at the angles of 40° and -40° with respect to the direction of the propagation of the electron (see Figure 12a). These parameters here are mainly chosen for minimizing the simulation efforts (simulation domain and time). However the physical principles behind the Kapitza-Dirac effect can be simply generalized to different wavelengths. The electron has a broadening of only 2 nm in both longitudinal and transversal directions, and the problem is solved in two-dimensions. The two Gaussian optical beam, form a standing wave pattern along the *y*-axis, whereas along the *x*-axis they interfere as a propagating wave. Obviously, the diffraction of the electron beam by the standing wave pattern is fully pictured with the full wave analysis by directly solving eq. (8). In contrast, solutions based on SVA (eq. (11)) cannot reproduce the Kapitza – Dirac diffraction pattern (compare Figures 12 b and c).

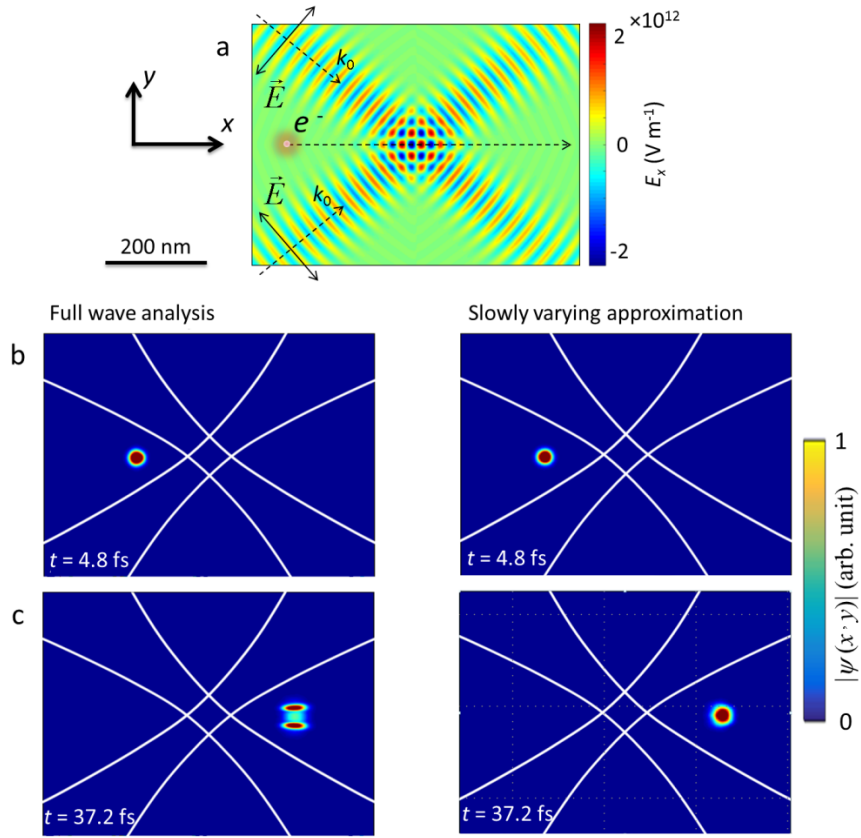

**Figure 12.** Comparison between full wave analysis and SVA for the Kpitza-Dirac effect. (a) The system comprises two Gaussian optical waves at a wavelength of 30 nm and an electron at a velocity of $\vec{v}_e = 0.02 c \hat{x}$ crossing the focal point of the optical beams. Electron wave function amplitude (initially Gaussian and with a broadening of 2 nm) at (b) $t = 4.8$ fs and (c) $t = 37.2$ fs, calculated using full wave analysis (left panels) and SVA approximation (right panels).

Indeed directly addressing eq. (8), though analytically challenging, would cast more accurate results, including the correct modulations of phase and amplitude of the electron wave function. Here, we outline our first attempts towards numerical realization of a Maxwell-Schrödinger self-consistent field approximation.

### 5.2. *Full wave analysis and self-consistent approach*

Indeed the most well-known self-consistent approach is the Hartree approximation. Hartree's method is based on the simplification of the many-body Hamiltonian of the Schrödinger equation, as

$$\frac{-\hbar^2}{2m_0}\nabla^2\psi(\vec{r}) + U^{ion}(\vec{r})\psi_i(\vec{r}) + U^{el}(\vec{r})\psi_i(\vec{r}) = E\psi(\vec{r}) \tag{12}$$

in the energy-domain representation; $U^{ion}(\vec{r})$ is the potential of the ions given by

$$U^{ion}(\vec{r}) = -Ze^2 \sum_{\vec{R}} \frac{1}{4\pi\varepsilon_0 |\vec{r}-\vec{R}|} \tag{13}$$

and the electron-electron interaction potential is averaged over the density of the electron charges as

$$U^{el}(\vec{r}) = -e\int dr' \frac{\rho(\vec{r}')}{4\pi\varepsilon_0 |\vec{r}-\vec{r}'|} \tag{14}$$

where the charge density distribution would be

$$\rho(\vec{r}) = -e\sum_{i}|\psi_i(\vec{r})|^2 \qquad (15)$$

and the sum is over all occupied one-electron levels. The Hartree approximation hence fails to consider the effect of the shape of the other electron wave functions over the single-electron wave function of interest, as only an averaging over the charge distributions is implied. In other words, it is only the averaged density which appears in eq. (15), regardless of the shape of the involved electronic orbitals. Despite the simplicity of the approximation, eq. (12) is still quite demanding from the numerical point of view. Additional improvements to the Hartree approximation are provided by the Slater determinant and the Hartree-Fock exchange potentials. Further introduction of self-consistent density-functional theory [48, 133, 134] and inclusion of pseudopotentials [135, 136] have added to the accuracy of single-electron approximations, albeit with an increasing level of numerical complexity. Moreover, generalization of the density-functional theory to the time-dependent Schrödinger equation has led to the introduction of the action integral, compared to the energy vibrational, which is to be minimized [137]. In practice however, electron-electron interactions in eq. (8), can be treated based on a time-dependent variation of the Hartree-Fock approximation [138].

*5.2.1 Self-consistent Maxwell-Schrödinger approximation*

Equations (12) to (15) should be solved self-consistently. The iteration includes insertion of initial approximations to the electron wave functions in the material under the investigation, computing potentials and insertion of the potentials into eq. (12), calculating the new set of wave functions, and repeating the cycle until $U^{el}(\vec{r})$ (or the wave functions) do not change any more.

Indeed, the solution of the $U^{el}(\vec{r})$ is often given by the Poisson equation ($\nabla^2 U^{el}(\vec{r}) = -e\rho(\vec{r})$), hence the problem is considered static. For swift electron beams in electron microscopes however, a common practice is to include an additional current density distribution given by [139]

$$\vec{J}_{el}(\vec{r},t) = \frac{e\hbar}{m_0} \text{Im}\left\{ \sum_i \psi_i(\vec{r},t) \vec{\nabla} \psi_i^*(\vec{r},t) \right\} - \frac{e^2 \rho(\vec{r},t)}{m_0} \vec{A}(\vec{r},t) \qquad (16)$$

This equation should be inserted into eq. (4a) to account for a self-consistent theory of field-approximation, including retardation. It is noticed that eq. (16) is necessary in order to satisfy the continuity of the electric charges, and especially important to model the electron-electron interactions similar to the Hartree approximation, though considering also screening effects. Particularly, from eq. (4a) one can obtain $\vec{\nabla} \cdot \vec{D}(\vec{r},t) = \rho(\vec{r},t)$, which in turn leads to the Poisson equation in the dielectric medium

$$\nabla^2 \varphi(\vec{r},t) = -\frac{\rho(\vec{r},t)}{\varepsilon_0 \varepsilon_r} \qquad (17)$$

whenever the Coulomb gauge is considered (accounting for Lorentz gauge leads to the Helmholtz equation for the scalar potential). $\varepsilon_r$ is the dielectric function of the material, or the environment surrounding the electron wave function. In vacuum the electron potential $U^{el}(\vec{r},t) = -e \int dr' \rho(\vec{r}',t)/4\pi\varepsilon_0 |\vec{r}-\vec{r}'|$ is simply reproduced, whereas in a material, inclusion of the dielectric function accounts for the screening effects [140]. In other words, introducing the current density distribution into the combined Maxwell-Schrödinger system of equations allows for the calculation of the magnetic vector potential- the latter is crucial for the self-consistent field theory including retardation picture.

We use the above mentioned retarded self-consistent field theory to simulate the dynamics of electron wave packets interacting with nanostructures and samples. Our crude assumption here is that the samples can be modeled using the dielectric theory, hence we do not account for the electron diffraction by the ionic potentials. Modeling of the latter effect is however achievable including appropriate pseudopotentials, but is not the subject of the present review. Our model however, considers collective electron-electron interactions representable by the dielectric function.

The following self–consistent steps form the current state of our developed Maxwell-Schrödinger numerical toolbox. We consider two different simulation domains of arbitrary sizes and arbitrary grids where the connections between them are held by accurately mapping the current distribution and potentials. In each time step we

- simulate the electron wave function by solving eq. (8), using a pseudospectral Fourier method [42, 141], which conserves the norm of the wave function,
- calculate the current-density distribution and project it from the Schrödinger domain into the Maxwell domain using an accurate mapping technique,
- solve the different field components using an FDTD algorithm,
- calculate the potentials from the field components by considering the Coulomb gauge theory, using $\nabla^2 \varphi = -\vec{\nabla} \cdot \vec{E}$ and $\partial \vec{A}/\partial t = -\vec{\nabla}\varphi - \vec{E}$, and
- map the potentials from the Maxwell domain into the Schrödinger domain.

Additionally, appropriate absorbing boundary conditions are satisfied for both the wave function and the field components in the Schrödinger and Maxwell domains, individually. It is finally noted here that for the sake of simplicity, the samples in present simulations included in this review are all modelled only with their bulk permittivity.

*5.2.2. Self – consistent treatment of dielectric laser accelerators*

Our first system includes a single slow-electron wave function interacting with a dielectric laser accelerator (DLA) [142, 143]. The structure of a DLA is composed of an optical grating, or a combination of those excited by laser beams at normal incidence configuration (see Figure 13). The phenomenon involved in the acceleration of electrons by the laser beam and the optical grating is the inverse Smith-Purcell effect. In order to facilitate the conservation of momentum and energy in the interaction of the laser beam with electrons, the same synchronization principle responsible for the Smith-Purcell radiation needs to be satisfied. We intend to benchmark the classical particle trajectory tracer algorithms with the quantum mechanical, though semi-classical, principles.

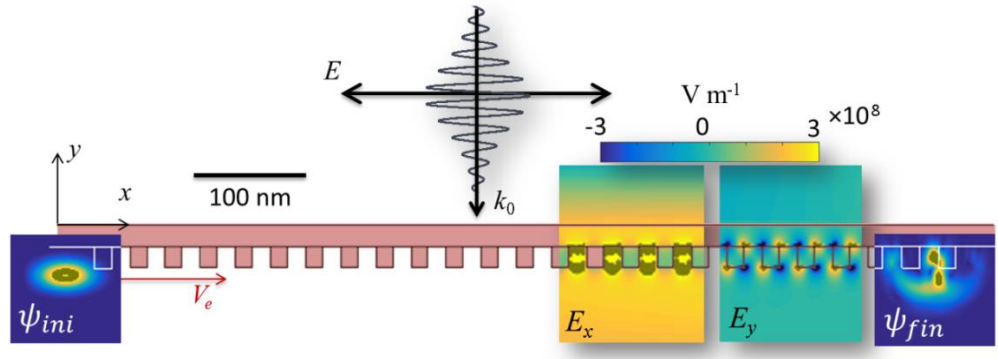

**Figure 13.** Inverse Smith–Purcell effect. (a) An electron pulse with longitudinal and transverse broadenings of 5 nm and 10 nm, respectively, travels at a distance of 4 nm parallel to a silicon grating. The silicon grating has a period of 33.2 nm and is illuminated with a laser pulse at a carrier wavelength of 830 nm and a temporal broadening of 80 fs. The spatial distribution of the electric field components at a given time is depicted in the inset. Snapshots of the spatial distribution of the initial and final electron wave function have been also shown [42].

A silicon grating with the period of 33.2 nm is considered which helps to satisfy the synchronicity of the near-field excitations with slow electron beams, for a laser excitation at the carrier wavelength of 830 nm and a temporal broadening of 80 fs. Interestingly, we notice that the electron wave packet is not accelerated, but in fact is decelerated. This is due to the fact that the near-field distribution of the grating supports concomitant excitation of electric-field components normal and parallel to the electron trajectory. The strong recoil that the electron experiences because of the electric field component normal to its trajectory causes a defocusing of the electron wave function. Additionally, we intend to maintain the synchronicity between the electron wave packet and the acceleration forces expressed on it. Because of the fast acceleration of the electron wave function within the first few periods of the field oscillation, this synchronization is better achieved by introducing a chirped grating [42], as shown in Figure 14. We notice however, that there exists a tradeoff between bunching of the initial electron wave function into sub-packets, and acceleration of the electron wave packet. In other words, two simultaneous phenomena are observed, namely *bunching* and *acceleration*. A control parameter for a dominant observation of each phenomenon is the longitudinal broadening of the electron wave function relative to the period of the structure. For longitudinal broadenings much smaller than the period of the grating, it is the acceleration which prevails the bunching effect, and vice versa. This effect is better understood by comparing the dynamics of electron wave packets at different broadenings interacting with a chirped optical grating (see Figure 14a and b). The momentum representation of the electron wave packet after its interaction with the grating clearly demonstrates the different mechanisms underlining the electron chirping, which clearly depends on the longitudinal broadening of the electron wave packet (Figure 14c).

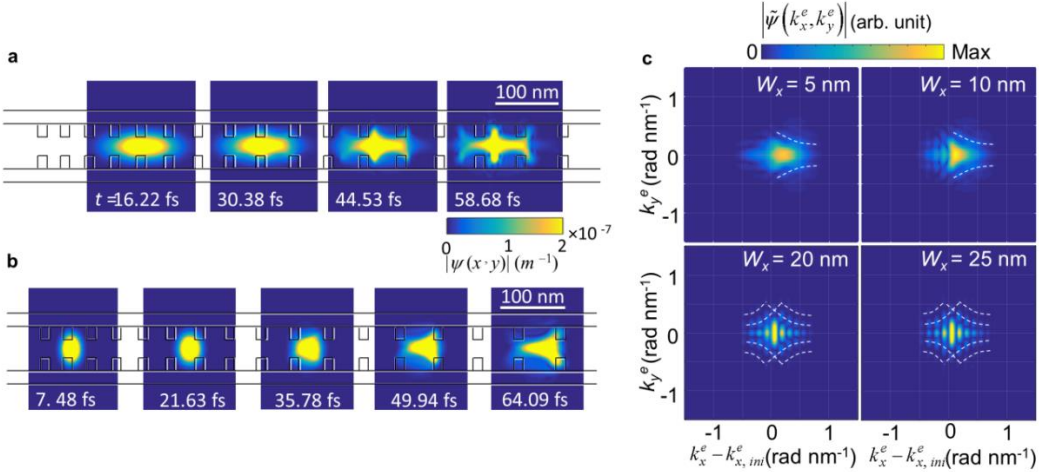

**Figure 14.** Interaction of an electron wavepacket at a velocity of 0.04*c* interacting with a chirped optical grating. The electron wavepacket has a transverse broadening of 5nm and a longitudinal broadening of (a) 20nm and (b) 5 nm. (c) The momentum representation of the wave packet with different longitudinal broadenings after interaction with the grating [42].

*5.2.2. Forming electron vortices by plasmon vortices*

Recent developments in electron microscopy involve triggering the shape of an electron wave function, to probe certain features of the sample excitation, to performing chiral dichroism [144, 145] or to decompose plasmonic excitations [70]. Among many possible shapes for the wave function, electron vortices have been extensively studied [145]. The attempts to propose appropriate electron vortices mainly depend on the design of holographic masks [146, 147] and magnetic monopoles [148]. Recently it was proposed, based on the conservation of the angular momentum, that free-space light which carries angular momentum may transfer the momentum to the electron waves during an elastic interaction [149]. We noticed however, that the free–space electron–photon interaction of this sort demands an amplitude for the light as large as $10^{12}$ V m$^{-1}$ to be able to observe a detectable transfer of momentum. The necessary large amplitude of the light is a perquisite to the Kapiza-Dirac effect. Moreover, light beams with angular momentum order (Laguerre-Gauss states) sustain a large domain with null intensity at the center, which increases in size by raising the angular momentum order. This effect puts limitations on an effective interaction of the electron waves with light, especially as the ponderomotive force pushes the electrons inside the regions with null intensity for the light.

Here, we propose at alternative approach which can be used to shape an initially Gaussian electron beam into a vortex beam, based on the formation of plasmon vortices [150]. We initially consider an aluminium disc with a thickness of only 10 nm, and a diameter of 400 nm, which is illuminated by a Laguerre-Gaussian optical beam [151] (with $l = 2$, and $p = 3$) at the carrier energy of 10 eV and temporal broadening of 2 fs. Diffraction of the light at the edges of the nanodisc excites surface plasmons which propagate towards the center of the disc and carry the same angular momentum as the excitation beam (see Figures 15 a and b). Thanks to the subwavelength propagation of surface plasmons [152], the excited plasmon waves propagate towards the center. The region with null intensity, which is intrinsic to the Laguerre-Gauss waves in free-space, is shrunk considerably in size, and an enhanced interaction with traveling electron wave packets is facilitated. To deliberate more on this idea, the interaction of a Gaussian electron wave packet at a kinetic energy of 10 keV which propagates towards the nanodisc along the $-z$ direction is simulated (Figure 15c to f). The electron wave packet has initial transverse and longitudinal broadenings of only 20 nm and 10 nm, respectively. The interaction of the electron wavepacket with vortex beams clearly shapes the electron wave packet. However, indeed we notice that the transformation of a single angular momentum order is not possible; rather we observe that the final electron wave packet is a complicated superposition of many angular momentum orders (see Figure 15e and h). For computing the angular momentum distribution of the final electron wave packet, a Fourier expansion versus the azimuthal angular orders as $\psi(x,y,z) = \sum_m \psi_m(\rho,z)\exp(im\varphi)$ was considered, where $\rho = \sqrt{x^2 + y^2}$ and $\varphi$ is the azimuthal angle. Furthermore, the angular momentum probability distribution (only in the azimuthal order) was calculated as $P_m = \int \rho\, d\rho\, dz\, |\psi_m(\rho,z)|^2$. In addition to the angular momentum, the linear momentum of the electron is also changed, both along the transverse direction (*xy*-plane) and along the longitudinal direction (*z*-plane), as shown in Figures 15e and 15f respectively. We attribute this behavior to the excitation of vortex surface plasmons and their complicated reflection of from the edges of the nanodisc. Moreover, the discrete energy levels of the electron, which are related to the discrete longitudinal momenta, are observed both as energy-loss and energy-gain peaks (see Figure 15f and g). These energy levels are not harmonics of the laser energy, in contrast with the observations in the PINEM experiments. We interestingly observe that even sub-harmonic energy peaks are also possible, thanks to the complicated nature of the laser excitation and the low-energy electron beam which is used here in the simulations. We attribute this behavior to the interference between plasmon excitations and laser-induced excitations, which will be discussed in more detail elsewhere. Moreover, it should be state that the above mentioned proposal depends on parameters like the interaction time and the longitudinal broadening the electron pulses. Whereas the former parameter indicates the sharpness of the peak of the angular momentum distribution of the final electron wave packet, the latter affects the inelastic interaction and gain and loss channels. It is expected that electrons with shorter longitudinal broadenings follow the quiver motion of the nea-field in a more effective way, before being bunched because of inelastic interaction with the near-field distribution of the nano-disc.

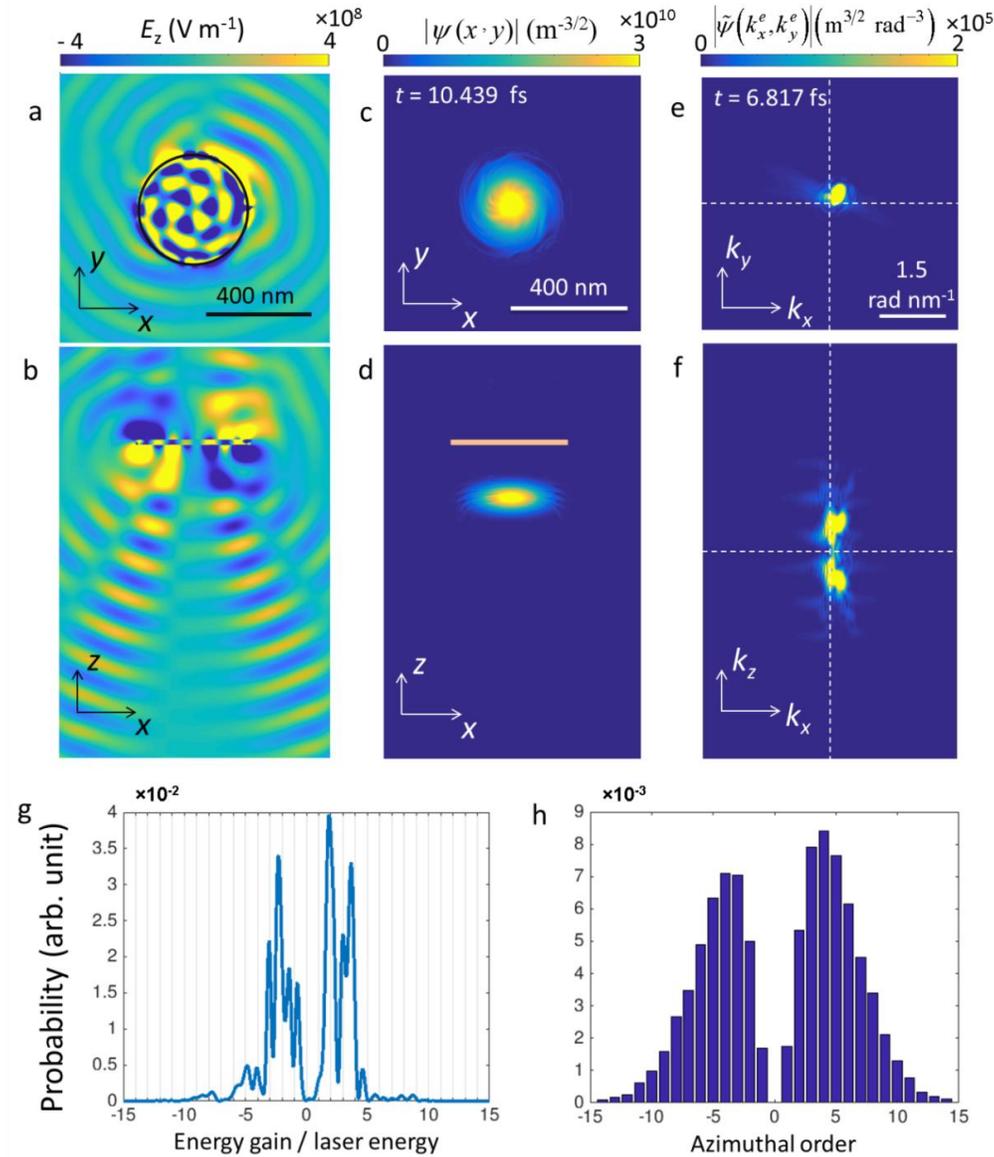

**Figure 15.** Interaction of an electron wave packet at a kinetic energy of 10 keV, with transverse and longitudinal broadenings of 50 nm and 20 nm respectively. The wavepacket interacts with vortex surface plasmons propagating in an aluminum nanodisc. The *z*-component of the laser-induced electric field in the (a) *xy*-plane recorded 5 nm above the structure, and (b) *xz*-plane passing through the middle of the disc. Spatial distribution of the electron wave function after the interaction, in the (c) *xy*-plane passing through the center of the wave function, and (d) *yz*-plane passing through the middle of the nanodisc. Final electron wave function in the momentum space, in the (e) $k_x - k_y$ plane at $k_z = 0$, and (f) in the $k_x - k_z$ plane at $k_y = 0$. (g) Energy distribution and (h) Angular momentum distribution of the final electron wave function.

## 6. Summary

Electron probes and optical excitations are both versatile tools for characterizing the sample excitations. Combination of both in recently developed techniques like PINEM and EEGS, has opened a plethora of interesting applications, proposing e.g. time-resolved spectroscopy, imaging, and diffraction. Concerning the intensity of the laser excitation, three distinguished domains are observed, namely electron-induced, electron-assisted, and photon-induced domains. Within the photon-assisted domain, one might observe interferences in the time-energy phase space which are recordable with EELS or CL techniques. However, we notice that such interference patterns demand a perfect synchronization between the electron and photon beams upon their arrival at the sample, or we phrase this as the crucial need for mutual coherence between the near-field of the electron and the laser beam. More or less all the mentioned domains are theoretically treated adiabatically, or by accepting non-recoil approximations. Here, we systematically studied specific cases where the adiabatic approximation break down, for which a self-consistent field theory would seem to be necessary. We proposed and developed two frameworks within the classical and quantum simulations. The former is achieved by combining the Maxwell and Lorentz equations in a self-consistent, time-domain, numerical toolbox, and was used to understand the classical recoil which the electron receives in its interaction with near-field distributions. The latter quantum framework is developed by combining the Maxwell and Schrödinger equations in a time-domain self–consistent numerical method, and was used to investigate the dynamics of slow-electron wavepackets propagating through the electromagnetic field distributions and nanostructures. We observe diffraction patterns and non-symmetric loss and gain patterns, which are not describable with analytical approximations. The applications of the proposed framework spans acceleration, point-projection electron microscopy, photoemission, and Kapitza-Dirac effect. More important however is, to combine the numerical experimentations with analytical understandings, to either better understand the observations, or to improve the adiabatic approximations. Within the outlines of the abovementioned classifications, an interesting approach to be taken in the electron science will be to control electron-light-matter interaction cycle with both shaped electrons and shaped light. Of particular interest in this direction is to study the dynamics of charge density waves to investigate emergent quantum behaviors and strongly correlated electron systems [153]. Whereas the momentum-space engineering of the associated macroscopic excitations will be achieved by the polarization states of the light, shaped electron can be used to symmetrically decompose the phase and amplitude oscillations, to distinguish Higgs modes as an example [154]. Within the concept of spontaneous symmetry breaking and at the intermediate level of the coupling strength, the rotationally symmetric effective potential of the system might impose two different oscillation modes, so called Goldstone mode and amplitude mode (Higgs mode). As the latter is distinguished by the oscillations along the radial direction, an electron probe, which in contrast to the vortex electrons imposes a radial symmetry, might be used to decompose the Higgs and Golstone oscillations.

**Acknowledgement**

NT gratefully thanks Christoph Lienau from Carl von Ossietzky Universität Oldenburg, Wilfried Sigle and Peter van Aken from Stuttgart Center for Electron Microscopy for fruitful discussions and joint investigations.

**References**

[1]   H. Batelaan, "Colloquium: Illuminating the Kapitza-Dirac effect with electron matter optics," Rev. Mod. Phys. 79 (2007) p. 929-941.

[2]   F. Ehlotzky, K. Krajewska, and J. Z. Kaminski, "Fundamental processes of quantum electrodynamics in laser fields of relativistic power*,"* Rep. Prog. Phys. 72 (2009) 046401.

[3]   V. Tsakanov and H. Wiedemann, *Brilliant Light in Life and Material Sciences*. Springer, The Netherlands, 2007.

[4]   C. Pellegrini, A. Marinelli, and S. Reiche, "The physics of x-ray free-electron lasers," Rev. Mod. Phys. 88 (2016) 015006.

[5]   P. R. Ribic and G. Margaritondo, "The Physics of X-Ray Free Electron Lasers (X-Fels): An Elementary Approach," Epioptics-11: Proceedings of the 49th Course of the International School of Solid State Physics 32 (2012) p. 68-91.

[6]   B. W. J. McNeil and N. R. Thompson, "X-ray free-electron lasers," Nat. Photonics 4 (2010) p. 814-821.

[7]   G. R. Blumenthal and R. J. Gould, "Bremsstrahlung, Synchrotron Radiation, and Compton Scattering of High-Energy Electrons Traversing Dilute Gases," Rev. Mod. Phys. 42 (1970) p. 237.

[8]   B. J. Siwick, J. R. Dwyer, R. E. Jordan, and R. J. D. Miller, "An atomic-level view of melting using femtosecond electron diffraction," Science 302 (2003) p. 1382-85.

[9]   B. Barwick, D. J. Flannigan, and A. H. Zewail, "Photon-induced near-field electron microscopy," Nature 462 (2009) p. 902-906.

[10]  A. H. Zewail and J. M. Thomas, *4D Electron Microscopy, Imaging in Space and Time*. Singapore: Imperial College Press, Singapore, 2010.

[11]  A. H. Zewail, "Four-Dimensional Electron Microscopy," Science 328 (2010) p. 187-193.

[12]  S. T. Park, M. M. Lin, and A. H. Zewail, "Photon-induced near-field electron microscopy (PINEM): theoretical and experimental," New J. Phys.12 (2010) 123028.

[13]  G. Sciaini and R. J. D. Miller, "Femtosecond electron diffraction: heralding the era of atomically resolved dynamics," Rep. Prog. Phys. 74 (2011) 096101.

[14]  R. J. D. Miller, "Mapping Atomic Motions with Ultrabright Electrons: The Chemists' Gedanken Experiment Enters the Lab Frame," Ann. Rev. Phys. Chem. 65 (2014) pp. 583-604.

[15]  G. M. Vanacore, A. W. P. Fitzpatrick, and A. H. Zewail, "Four-dimensional electron microscopy: Ultrafast imaging, diffraction and spectroscopy in materials science and biology," Nano Today 11 (2016) p. 228-249.

[16]  Y. M. Lee, Y. J. Kim, Y. J. Kim, and O. H. Kwon, "Ultrafast electron microscopy integrated with a direct electron detection camera," Structural Dynamics 4 (2017) 044023.

[17]  A. Feist, K. E. Echternkamp, J. Schauss, S. V. Yalunin, S. Schafer, and C. Ropers, "Quantum coherent optical phase modulation in an ultrafast transmission electron microscope," *Nature* 521 (2015) p. 200.

[18]  L. Piazza, T. T. A. Lummen, E. Quinonez, Y. Murooka, B. W. Reed, B. Barwick*, et al.*, "Simultaneous observation of the quantization and the interference pattern of a plasmonic near-field," Nature Commun. 6 (2015) 6407.

[19]  K. Bücker, M. Picher, O. Crégut, T. LaGrange, B. Reed, S. T. Park*, et al.*, "Ultrafast transmission electron microscopy reveals electron dynamics and trajectories in a thermionic gun setup," in *European Microscopy Congress 2016: Proceedings*, ed: Wiley-VCH Verlag GmbH & Co. KGaA, 2016.

[20] G. L. Cao, S. S. Sun, Z. W. Li, H. F. Tian, H. X. Yang, and J. Q. Li, "Clocking the anisotropic lattice dynamics of multi-walled carbon nanotubes by four-dimensional ultrafast transmission electron microscopy," Sci. Rep. 5 (2015) 8404.

[21] D. Ehberger, J. Hammer, M. Eisele, M. Kruger, J. Noe, A. Hogele, *et al.*, "Highly Coherent Electron Beam from a Laser-Triggered Tungsten Needle Tip," Phys. Rev. Lett. 114 (2015) 227601.

[22] M. Walbran, A. Gliserin, K. Jung, J. Kim, and P. Baum, "5-Femtosecond Laser-Electron Synchronization for Pump-Probe Crystallography and Diffraction," *Phys. Rev. Appl.* 4 (2015) 044013.

[23] W. Verhoeven, J. F. M. van Rens, E. R. Kieft, P. H. A. Mutsaers, and O. J. Luiten, "High quality ultrafast transmission electron microscopy using resonant microwave cavities," *Ultramicroscopy* 188 (2018) p. 85-89.

[24] W. Verhoeven, J. F. M. v. Rens, M. A. W. v. Ninhuijs, W. F. Toonen, E. R. Kieft, P. H. A. Mutsaers, *et al.*, "Time-of-flight electron energy loss spectroscopy using TM110 deflection cavities," Structural Dynamics 3 (2016) p. 054303.

[25] J. Maxson, D. Cesar, G. Calmasini, A. Ody, P. Musumeci, and D. Alesini, "Direct Measurement of Sub-10 fs Relativistic Electron Beams with Ultralow Emittance," Phys. Rev. Lett. 118 (2017) p. 154802.

[26] J. Yang, M. Guehr, T. Vecchione, M. S. Robinson, R. Li, N. Hartmann, *et al.*, "Diffractive imaging of a rotational wavepacket in nitrogen molecules with femtosecond megaelectronvolt electron pulses," Nature Commun. 7 (2016) p. 11232.

[27] S. P. Weathersby, G. Brown, M. Centurion, T. F. Chase, R. Coffee, J. Corbett, *et al.*, "Mega-electron-volt ultrafast electron diffraction at SLAC National Accelerator Laboratory," Rev. Sci. Instrum. 86 (2015) p. 073702.

[28] K. E. Priebe, C. Rathje, S. V. Yalunin, T. Hohage, A. Feist, S. Schaer, *et al.*, "Attosecond electron pulse trains and quantum state reconstruction in ultrafast transmission electron microscopy," Nature Photon. 11 (2017) p. 793.

[29] P. Baum, "On the physics of ultrashort single-electron pulses for time-resolved microscopy and diffraction," Chem. Phys. 423 (2013) p. 55-61.

[30] A. Gliserin, A. Apolonski, F. Krausz, and P. Baum, "Compression of single-electron pulses with a microwave cavity," New J. Phys. 14 (2012) 073055.

[31] C. Kealhofer, W. Schneider, D. Ehberger, A. Ryabov, F. Krausz, and P. Baum, "All-optical control and metrology of electron pulses," Science 352 (2016) p. 429-433.

[32] T. C. Weinacht, J. Ahn, and P. H. Bucksbaum, "Controlling the shape of a quantum wavefunction," Nature 397 (1999) p. 233-235.

[33] C. Brif, R. Chakrabarti, and H. Rabitz, "Control of quantum phenomena: past, present and future," New J. Phys.12 (2010) 075008.

[34] N. Talebi, "Interaction of electron beams with optical nanostructures and metamaterials: from coherent photon sources towards shaping the wave function," J. Opt. (Bristol, U. K.) 19 (2017) p. 103001.

[35] N. Talebi, "Spectral Interferometry with Electron Microscopes," Sci. Rep. 6 (2016) 33874.

[36] H. W. Fink, W. Stocker, and H. Schmid, "Holography with Low-Energy Electrons," Phys. Rev. Lett. 65 (1990) p. 1204-1206.

[37] T. Latychevskaia, J. N. Longchamp, C. Escher, and H. W. Fink, "Holography and coherent diffraction with low-energy electrons: A route towards structural biology at the single molecule level," Ultramicroscopy 159 (2015) p. 395-402.

[38] P. Gross, M. Esmann, S. F. Becker, J. Vogelsang, N. Talebi, and C. Lienau, "Plasmonic nanofocusing - grey holes for light," Advances in Physics-X 1 (2016) p. 297-330.

[39] M. Muller, V. Kravtsov, A. Paarmann, M. B. Raschke, and R. Ernstorfer, "Nanofocused Plasmon-Driven Sub-10 fs Electron Point Source," Acs Photonics 3 (2016) p. 611-619.

[40] J. Vogelsang, J. Robin, B. J. Nagy, P. Dombi, D. Rosenkranz, M. Schiek, *et al.*, "Ultrafast Electron Emission from a Sharp Metal Nanotaper Driven by Adiabatic Nanofocusing of Surface Plasmons," Nano Lett. 15 (2015) p. 4685-4691.

[41] M. Kozak, T. Eckstein, N. Schonenberger, and P. Hommelhoff, "Inelastic ponderomotive scattering of electrons at a high-intensity optical travelling wave in vacuum," Nature Physics 14 (2018) p. 121–125.

[42] N. Talebi, "Schrödinger electrons interacting with optical gratings: quantum mechanical study of the inverse Smith-Purcell effect," New J. Phys., 18 (2016) 123006.

[43] D. Y. Sergeeva, A. P. Potylitsyn, A. A. Tishchenko, and M. N. Strikhanov, "Smith-Purcell radiation from periodic beams," Optics Express 25 (2017) p. 26310-26328.

[44] S. Tsesses, G. Bartal, and I. Kaminer, "Light generation via quantum interaction of electrons with periodic nanostructures," Phys. Rev. A 95 ( 2017) 013832.

[45] I. Kaminer, M. Mutzafi, A. Levy, G. Harari, H. H. Sheinfux, S. Skirlo, *et al.*, "Quantum Cerenkov Radiation: Spectral Cutoffs and the Role of Spin and Orbital Angular Momentum," Phys. Rev. X 6 (2016) 011006.

[46] J. Peatross, C. Muller, K. Z. Hatsagortsyan, and C. H. Keitel, "Photoemission of a single-electron wave packet in a strong laser field," Phys. Rev. Lett. 100 (2008) 153601.

[47] A. J. White, M. Sukharev, and M. Galperin, "Molecular nanoplasmonics: Self-consistent electrodynamics in current-carrying junctions," Phys. Rev. B 86 (2012) p. 205324.

[48] W. Kohn and L. J. Sham, "Self-Consistent Equations Including Exchange and Correlation Effects," Phys. Rev. 140 (1965) p. 1133.

[49] R. F. Egerton, "Electron energy-loss spectroscopy in the TEM," Rep. Prog. Phys. 72 (2009) 016502.

[50] F. J. García de Abajo, "Optical excitations in electron microscopy," Rev. Mod. Phys. 82 (2010) p. 209-275.

[51] V. J. Keast, A. J. Scott, R. Brydson, D. B. Williams, and J. Bruley, "Electron energy-loss near-edge structure – a tool for the investigation of electronic structure on the nanometre scale," Journal of Microscopy 203 (2001) p. 135-175.

[52] T. Coenen and N. M. Haegel, "Cathodoluminescence for the 21st century: Learning more from light," Applied Physics Reviews 4 (2017) 4985767.

[53] T. Coenen, S. V. den Hoedt, and A. Polman, "A New Cathodoluminescence System for Nanoscale Optics, Materials Science, and Geology," Microscopy Today 24 (2016) p. 12-19.

[54] M. Kociak and O. Stephan, "Mapping plasmons at the nanometer scale in an electron microscope," Chem. Soc. Rev. 43 (2014) p. 3865-3883.

[55] Z. Mahfoud, A. T. Dijksman, C. Javaux, P. Bassoul, A. L. Baudrion, J. Plain, *et al.*, "Cathodoluminescence in a Scanning Transmission Electron Microscope: A Nanometer-Scale Counterpart of Photoluminescence for the Study of II-VI Quantum Dots," J. Phys. Chem. Lett. 4 (2013) p. 4090-4094.

[56]  N. Yamamoto, "Development of high-resolution cathodoluminescence system for STEM and application to plasmonic nanostructures," Microscopy 65 (2016) p. 282-295.

[57]  B. J. M. Brenny, T. Coenen, and A. Polman, "Quantifying coherent and incoherent cathodoluminescence in semiconductors and metals," Journal of Applied Physics 115, (2014) 244307.

[58]  R. H. Ritchie and A. Howie, "Inelastic-Scattering Probabilities in Scanning-Transmission Electron-Microscopy," Philos. Mag. A 58 (1988) p. 753-767.

[59]  F. J. García de Abajo and M. Kociak, "Probing the Photonic Local Density of States with Electron Energy Loss Spectroscopy," Phys. Rev. Lett. 100 (2008) p. 106804.

[60]  P. Schattschneider, *Fundamentals of Inelastic Electron Scattering*. Wien, Springer-Verlag, New York,1986.

[61]  N. Talebi, "A directional, ultrafast and integrated few-photon source utilizing the interaction of electron beams and plasmonic nanoantennas," New J. Phys.16 (2014) 053021.

[62]  A. Losquin and M. Kociak, "Link between Cathodoluminescence and Electron Energy Loss Spectroscopy and the Radiative and Full Electromagnetic Local Density of States," Acs Photonics 2 (2015) p. 1619-1627.

[63]  A. Losquin, L. F. Zagonel, V. Myroshnychenko, B. Rodriguez-Gonzalez, M. Tence, L. Scarabelli*, et al.*, "Unveiling Nanometer Scale Extinction and Scattering Phenomena through Combined Electron Energy Loss Spectroscopy and Cathodoluminescence Measurements," Nano Letters 15 (2015) p. 1229-1237.

[64]  M. Kociak, O. Stephan, A. Gloter, L. F. Zagonel, L. H. G. Tizei, M. Tence*, et al.*, "Seeing and measuring in colours: Electron microscopy and spectroscopies applied to nano-optics," Comptes Rendus Physique 15 (2014) p. 158-175.

[65]  J. Nelayah, M. Kociak, O. Stephan, F. J. G. de Abajo, M. Tence, L. Henrard*, et al.*, "Mapping surface plasmons on a single metallic nanoparticle," Nature Physics 3 (2007) p. 348-353.

[66]  E. P. Bellido, Y. Zhang, A. Manjavacas, P. Nordlander, and G. A. Botton, "Plasmonic Coupling of Multipolar Edge Modes and the Formation of Gap Modes," Acs Photonics 4 (2017) p. 1558-1565.

[67]  A. Campos, A. Arbouet, J. Martin, D. Gerard, J. Proust, J. Plain*, et al.*, "Plasmonic Breathing and Edge Modes in Aluminum Nanotriangles," Acs Photonics 4 (2017) p. 1257-1263.

[68]  Y. Fujiyoshi, T. Nemoto, and H. Kurata, "Studying substrate effects on localized surface plasmons in an individual silver nanoparticle using electron energy-loss spectroscopy," Ultramicroscopy 175 (2017) p. 116-120.

[69]  L. Gu, W. Sigle, C. T. Koch, B. Ogut, P. A. van Aken, N. Talebi*, et al.*, "Resonant wedge-plasmon modes in single-crystalline gold nanoplatelets," Physical Review B 83 (2011) 195433.

[70]  G. Guzzinati, A. Beche, H. Lourenco-Martins, J. Martin, M. Kociak, and J. Verbeeck, "Probing the symmetry of the potential of localized surface plasmon resonances with phase-shaped electron beams," Nature Commun. 8 (2017) 14999.

[71]  B. Ogut, N. Talebi, R. Vogelgesang, W. Sigle, and P. A. van Aken, "Toroidal Plasmonic Eigenmodes in Oligomer Nanocavities for the Visible," Nano Letters 12 (2012) p. 5239-5244.

[72]  D. Rossouw and G. A. Botton, "Plasmonic Response of Bent Silver Nanowires for Nanophotonic Subwavelength Waveguiding," Phys. Rev. Lett. 110 (2013) 23432286.
[73]  P. Shekhar, M. Malac, V. Gaind, N. Dalili, A. Meldrum, and Z. Jacob, "Momentum-Resolved Electron Energy Loss Spectroscopy for Mapping the Photonic Density of States," Acs Photonics 4 (2017) p. 1009-1014.
[74]  N. Talebi, C. Ozsoy-Keskinbora, H. M. Benia, K. Kern, C. T. Koch, and P. A. van Aken, "Wedge Dyakonov Waves and Dyakonov Plasmons in Topological Insulator Bi2Se3 Probed by Electron Beams," Acs Nano 10 (2016) p. 6988-6994.
[75]  N. Talebi, W. Sigle, R. Vogelgesang, M. Esmann, S. F. Becker, C. Lienau, *et al.*, "Excitation of Mesoscopic Plasmonic Tapers by Relativistic Electrons: Phase Matching versus Eigenmode Resonances," Acs Nano 9 (2015) p. 7641-7648.
[76]  Y. Zhu, P. N. H. Nakashima, A. M. Funston, L. Bourgeois, and J. Etheridge "Topologically Enclosed Aluminum Voids as Plasmonic Nanostructures," Acs Photonics 11 (2017) p. 11383–11392
[77]  B. Ogut, R. Vogelgesang, W. Sigle, N. Talebi, C. T. Koch, and P. A. van Aken, "Hybridized Metal Slit Eigenmodes as an Illustration of Babinet's Principle," Acs Nano 5 (2011) p. 6701-6706.
[78]  M. Husnik, F. von Cube, S. Irsen, S. Linden, J. Niegemann, K. Busch, *et al.*, "Comparison of electron energy-loss and quantitative optical spectroscopy on individual optical gold antennas," Nanophotonics 2 (2013) p. 241-245.
[79]  A. L. Koh, K. Bao, I. Khan, W. E. Smith, G. Kothleitner, P. Nordlander, *et al.*, "Electron Energy-Loss Spectroscopy (EELS) of Surface Plasmons in Single Silver Nanoparticles and Dimers: Influence of Beam Damage and Mapping of Dark Modes," Acs Nano 3 (2009) p. 3015-3022.
[80]  B. Schroder, T. Weber, S. V. Yalunin, T. Kiel, C. Matyssek, M. Sivis, *et al.*, "Real-space imaging of nanotip plasmons using electron energy loss spectroscopy," Phys. Rev. B 92 (2015) 085411.
[81]  S. Guo, N. Talebi, W. Sigle, R. Vogelgesang, G. Richter, M. Esmann, *et al.*, "Reflection and Phase Matching in Plasmonic Gold Tapers," Nano Lett. 16 (2016) p. 6137-6144.
[82]  S. J. Smith and E. M. Purcell, "Visible Light from Localized Surface Charges Moving across a Grating," Phys. Rev. 92 (1953) p. 1069-1069.
[83]  A. Gover, P. Dvorkis, and U. Elisha, "Angular Radiation-Pattern of Smith-Purcell Radiation," J. Opt. Soc. Am. B 1 (1984) p. 723-728.
[84]  A. Gover and P. Sprangle, "A Unified Theory of Magnetic Bremsstrahlung, Electrostatic Bremsstrahlung, Compton-Raman Scattering, and Cerenkov-Smith-Purcell Free-Electron Lasers," IEEE J. Quantum Electron. 17 (1981) p. 1196-1215.
[85]  N. Yamamoto, F. J. G. de Abajo, and V. Myroshnychenko, "Interference of surface plasmons and Smith-Purcell emission probed by angle-resolved cathodoluminescence spectroscopy," Phys. Rev. B 91 (2015) 125144.
[86]  V. L. Ginzburg, "Transition Radiation and Transition Scattering," Physica Scripta T2 (1982) p. 182-191.
[87]  J. P. Verboncoeur, "Particle simulation of plasmas: review and advances," Plasma Physics and Controlled Fusion 47 (2005) p. A231-A260.
[88]  C. S. Meierbachtol, A. D. Greenwood, J. P. Verboncoeur, and B. Shanker, "Conformal Electromagnetic Particle in Cell: A Review," IEEE Trans. Plasma Sci. 43 (2015) p. 3778-3793.

[89] R. Lee, H. H. Klein, and J. P. Boris, "Computer-Simulation of Electromagnetic Instabilities in a Relativistic Plasma," Bull. Am. Phys. Soc. 18 (1973) p. 633-633.
[90] R. Lee, J. P. Boris, and I. Haber, "Electromagnetic Simulation Codes for Relativistic Plasmas," Bull. Am. Phys. Soc. 17 (1972) p. 1048-1048.
[91] R. A. Fonseca, J. Vieira, F. Fiuza, A. Davidson, F. S. Tsung, W. B. Mori, *et al.*, "Exploiting multi-scale parallelism for large scale numerical modelling of laser wakefield accelerators," Plasma Phys. Controlled Fusion 55 (2013) 124011.
[92] C. G. R. Geddes, C. Toth, J. Van Tilborg, E. Esarey, C. B. Schroeder, D. Bruhwiler, *et al.*, "Laser guiding at relativistic intensities and wakefield particle acceleration in plasma channels," AIP Conference Proceedings 737 (2004) p. 521.
[93] A. Pukhov, Z. M. Sheng, and J. Meyer-ter-Vehn, "Particle acceleration in relativistic laser channels," Physics of Plasmas 6 (199) p. 2847-2854.
[94] N. Talebi, M. Shahabadi, W. Khunsin, and R. Vogelgesang, "Plasmonic grating as a nonlinear converter-coupler," Optics Express 20 (2012) p. 1392-1405.
[95] N. Talebi and M. Shahabadi, "All-optical wavelength converter based on a heterogeneously integrated GaP on a silicon-on-insulator waveguide," Journal of the Optical Society of America B - Optical Physics 27 (2010) p. 2273-2278.
[96] S. N. Lyle, *Self force and inertia, Old light on new ideas*. Springer, Heidelberg: 2010.
[97] J. D. Jackson, *Classical Electrodynamics*. John Wiley & Sons, Inc., United States: 1999.
[98] O. L. Krivanek, T. C. Lovejoy, N. Dellby, T. Aoki, R. W. Carpenter, P. Rez, *et al.*, "Vibrational spectroscopy in the electron microscope," Nature 514 (2014) p. 209.
[99] M. J. Lagos, A. Trügler, U. Hohenester, and P. E. Batson, "Mapping vibrational surface and bulk modes in a single nanocube," Nature 543 (2017) p. 529.
[100] F. J. G. d. Abajo and M. Kociak, "Electron energy-gain spectroscopy," *New J. Phys.* 10 (2008) p. 073035.
[101] F. J. García de Abajo, B. Barwick, and F. Carbone, "Electron diffraction by plasmon waves," Phys. Rev. B 94 (2016) p. 041404.
[102] M. M. Nieto, "Diffraction of Electrons by Standing Electromagnetic Waves: The Kapitza-Dirac Effect," American Journal of Physics 37 (19690 p. 162-169.
[103] A. Howie, "Photon interactions for electron microscopy applications," Eur. Phys. J. Appl. Phys. 54 (2011) p. 33502.
[104] S. T. Park and A. H. Zewail, "Photon-Induced Near Field Electron Microscopy," Ultrafast Imaging and Spectroscopy 8845 (2013) ISBN: 9780819496959.
[105] A. H. Zewail, "4D ultrafast electron diffraction, crystallography, and microscopy," Annu. Rev. Phys. Chem. 57 (2006) p. 65-103.
[106] B. K. Yoo, Z. X. Su, J. M. Thomas, and A. H. Zewail, "On the dynamical nature of the active center in a single-site photocatalyst visualized by 4D ultrafast electron microscopy," Proceedings of the National Academy of Sciences of the United States of America 113 (2016) p. 503-508.
[107] B. W. Shore and J. H. Eberly, "Analytic Approximations in Multilevel Excitation Theory," Bull. Am. Phys. Soc. 23 (1978) p. 34-34.
[108] S. Aaronson and A. Arkhipov, "The Computational Complexity of Linear Optics," Stoc 11: Proceedings of the 43rd Acm Symposium on Theory of Computing (2011) p. 333-342.

[109] A. Crespi, R. Osellame, R. Ramponi, D. J. Brod, E. F. Galvao, N. Spagnolo, *et al.*, "Integrated multimode interferometers with arbitrary designs for photonic boson sampling," Nature Photonics 7 (2013) p. 545-549.

[110] K. E. Echternkamp, A. Feist, S. Schafer, and C. Ropers, "Ramsey-type phase control of free-electron beams," Nature Physics 12 (2016) p. 1000.

[111] Y. Morimoto and P. Baum, "Diffraction and microscopy with attosecond electron pulse trains," Nature Physics 14 (2018) p. 252.

[112] A. Losquin and T. T. A. Lummen, "Electron microscopy methods for space-, energy-, and time-resolved plasmonics," Frontiers of Physics 12 (2017) p. 127301.

[113] A. Ryabov and P. Baum, "Electron microscopy of electromagnetic waveforms," Science 353 (2016) p. 374-377.

[114] A. R. Bainbridge, C. W. B. Myers, and W. A. Bryan, "Femtosecond few- to single-electron point-projection microscopy for nanoscale dynamic imaging," Structural Dynamics 3 (2016) 023612.

[115] E. Quinonez, J. Handali, and B. Barwick, "Femtosecond photoelectron point projection microscope," Review of Scientific Instruments 84 (2013) 103710.

[116] D. Gabor, "A New Microscopic Principle," Nature 161 (1948) p. 777-778.

[117] A. Feist, N. Bach, N. R. da Silva, T. Danz, M. Moller, K. E. Priebe, *et al.*, "Ultrafast transmission electron microscopy using a laser-driven field emitter: Femtosecond resolution with a high coherence electron beam," *Ultramicroscopy,* vol. 176, pp. 63-73, May 2017.

[118] C. J. Powell and J. B. Swan, "Origin of the Characteristic Electron Energy Losses in Aluminum," Phys. Rev. 115 (1959) p. 869-875.

[119] C. J. Powell and J. B. Swan, "Origin of the Characteristic Electron Energy Losses in Magnesium," Physical Review 116 (1959) p. 81-83.

[120] R. Garciamolina, A. Grasmarti, A. Howie, and R. H. Ritchie, "Retardation Effects in the Interaction of Charged-Particle Beams with Bounded Condensed Media," Journal of Physics C-Solid State Physics 18 (1985) p. 5335-5345.

[121] N. Talebi, W. Sigle, R. Vogelgesang, and P. van Aken, "Numerical simulations of interference effects in photon-assisted electron energy-loss spectroscopy," New J. Phys. 15 (2013) 053013.

[122] A. Howie, "Stimulated excitation electron microscopy and spectroscopy," Ultramicroscopy 151 (2015) p. 116-121.

[123] A. Poddubny, I. Iorsh, P. Belov, and Y. Kivshar, "Hyperbolic metamaterials," Nature Photonics 7 (2013) p. 948-957.

[124] F. Roder and H. Lichte, "Inelastic electron holography - first results with surface plasmons," European Physical Journal-Applied Physics 54 (2011) 33504.

[125] P. Schattschneider and H. Lichte, "Correlation and the density-matrix approach to inelastic electron holography in solid state plasmas," Physical Review B 71 (2005) 045130.

[126] F. P. Schmidt, H. Ditlbacher, U. Hohenester, A. Hohenau, F. Hofer, and J. R. Krenn, "Dark Plasmonic Breathing Modes in Silver Nanodisks," Nano Letters 12 (2012) p. 5780-5783.

[127] O. Smirnova, M. Spanner, and M. Ivanov, "Analytical solutions for strong field-driven atomic and molecular one- and two-electron continua and applications to strong-field problems," Phys. Rev. A 77 (2008) 033407.

[128] D. M. Wolkow, "On a mass of solutions of the Dirac equation.," Zeitschrift Fur Physik 94 (1935) p. 250-260.
[129] E. Kasper, "Generalization of Schrodingers Wave Mechanics for Relativistic Regions of Validity," Zeitschrift Für Naturforschung Section A A-28 (1973) p. 216-221.
[130] S. T. Park, "Propagation of a relativistic electron wave packet in the Dirac equation," Physical Review A 86 (2012) 062105.
[131] F. J. G. de Abajo, A. Asenjo-Garcia, and M. Kociak, "Multiphoton Absorption and Emission by Interaction of Swift Electrons with Evanescent Light Fields," Nano Letters 10 (2010) p. 1859-1863.
[132] D. Wolf, L. A. Rodriguez, A. Beche, E. Javon, L. Serrano, C. Magen, *et al.*, "3D Magnetic Induction Maps of Nanoscale Materials Revealed by Electron Holographic Tomography," Chemistry of Materials 27 (2015) p. 6771-6778.
[133] R. O. Jones and O. Gunnarsson, "The Density Functional Formalism, Its Applications and Prospects," *Rev. Mod. Phys.* 61 (1989) p. 689-746.
[134] E. J. Baerends, "Perspective on "Self-consistent equations including exchange and correlation effects" - Kohn W, Sham LJ (1965) Phys Rev A 140 : 133-1138," *Theoretical Chemistry Accounts,* 103 (2000) p. 265-269.
[135] B. Walker and R. Gebauer, "Ultrasoft pseudopotentials in time-dependent density-functional theory," J. Chem. Phys. 127 (2007) p. 164106.
[136] J. Harris and R. O. Jones, "Pseudopotentials in Density-Functional Theory," Phys.Rev. Lett. 41 (1978) p. 191-194.
[137] E. Runge and E. K. U. Gross, "Density-Functional Theory for Time-Dependent Systems," Phys. Rev. Lett. 52 (1984) p. 997-1000.
[138] X. S. Li, S. M. Smith, A. N. Markevitch, D. A. Romanov, R. J. Levis, and H. B. Schlegel, "A time-dependent Hartree-Fock approach for studying the electronic optical response of molecules in intense fields," Physical Chemistry Chemical Physics 7 (2005) p. 233-239.
[139] P. W. Hawkes and K. E., *Principles of Electron Optics* vol. 3. Academic Press, London,1996.
[140] N. W. Ashcroft and N. D. Mermin, *Solid State Physics.* Thomson Learning, Inc., USA, 1976.
[141] H. Tal-Ezer and R. Kosloff, "An accurate and efficient scheme for propagating the time dependent Schrödinger equation," The Journal of Chemical Physics 81 (1984) p. 3967-3971.
[142] J. Breuer, J. McNeur, and P. Hommelhoff, "Dielectric laser acceleration of electrons in the vicinity of single and double grating structures-theory and simulations," J. Phys. B: At. Mol. Phys. 47 (2014) 234004.
[143] J. Breuer and P. Hommelhoff, "Laser-Based Acceleration of Nonrelativistic Electrons at a Dielectric Structure," Phys. Rev. Lett. 111 (2013) 134803.
[144] P. Schattschneider, S. Löffler, M. Stöger-Pollach, and J. Verbeeck, "Is magnetic chiral dichroism feasible with electron vortices?," Ultramicroscopy 136 (2014) p. 81-85.
[145] S. M. Lloyd, M. Babiker, G. Thirunavukkarasu, and J. Yuan, "Electron vortices: Beams with orbital angular momentum," Rev. Mod. Phys. 89 (2017) p. 035004.
[146] D. Pohl, S. Schneider, P. Zeiger, J. Rusz, P. Tiemeijer, S. Lazar, *et al.*, "Atom size electron vortex beams with selectable orbital angular momentum," Sci. Rep. 7 (2017) p. 934.

[147] P. Schattschneider and J. Verbeeck, "Theory of free electron vortices," Ultramicroscopy 111 (2011) p. 1461-1468.

[148] A. Béché, R. Van Boxem, G. Van Tendeloo, and J. Verbeeck, "Magnetic monopole field exposed by electrons," Nature Physics 10 (2013) p. 26.

[149] J. Handali, P. Shakya, and B. Barwick, "Creating electron vortex beams with light," Optics Express 23 (2015) p. 5236-5243.

[150] G. Spektor, D. Kilbane, K. Mahro, B. Frank, S. Ristok, L. Gal, *et al.*, "Revealing the subfemtosecond dynamics of orbital angular momentum in nanoplasmonic vortices," Science 355 (2017) p. 1187-1191.

[151] J. C. Gutierrez-Vega, "Fractionalization of optical beams: II. Elegant Laguerre-Gaussian modes," Optics Express 15 (2007) p. 6300-6313.

[152] M. I. Stockman, "Nanoplasmonics: past, present, and glimpse into future," Optics Express 19, (2011) p. 22029-22106.

[153] P. Coleman and A. J. Schofield, "Quantum criticality," Nature 433 (2005) p. 226-229.

[154] Y. Kuno, K. Suzuki, and I. Ichinose, "Effective Field Theory for Two-Species Bosons in an Optical Lattice: Multiple Order, the Nambu-Goldstone Bosons, the Higgs Mode, and Vortex Lattice," J. Phys. Soc. Jpn. 82 (2013) 124501.